\documentclass[twocolumn]{article}
\usepackage[disable]{todonotes}
\usepackage{geometry}
\usepackage[utf8]{inputenc}
\usepackage{authblk}
\usepackage{setspace}
\usepackage{graphicx}
\usepackage{subcaption}
\usepackage{amsmath}
\usepackage{lineno}
\usepackage{algorithm}
\usepackage{algorithmicx}
\usepackage[noend]{algpseudocode}
\usepackage{color}      
\usepackage[final]{changes} 

\usepackage{listings}
\definecolor{dkgreen}{rgb}{0,0.5,0.7}
\definecolor{gray}{rgb}{0.5,0.5,0.5}
\definecolor{mauve}{rgb}{0.58,0,0.82}
\floatname{algorithm}{Algorithm}

\usepackage[numbers,compress]{natbib}
\usepackage{hyperref}
\hypersetup{pdfborder=0 0 0}

\title{A Resource-Virtualized and Hardware-Aware Quantum Compilation Framework for Real Quantum Computing Processors}

\author[1,2,3*]{Hong-Ze Xu} 
\author[1,3]{Xu-Dan Chai}
\author[1,3*]{Meng-Jun Hu} 
\author[1,3]{Zheng-An Wang}
\author[1,3]{Yu-Long Feng}
\author[1,4]{Yu Chen}
\author[1,5]{Xinpeng Zhang}
\author[1,3]{Jingbo Wang}
\author[1,3]{Wei-Feng Zhuang}
\author[1,3]{Yu-Xin Jin}
\author[1,3]{Yirong Jin}
\author[1,3]{Haifeng Yu}
\author[1,6,7,8,9]{Heng Fan}
\author[1,2,3,9,10*]{Dong E. Liu}

\affil[1]{Beijing Academy of Quantum Information Sciences, Beijing 100193, China}
\affil[2]{State Key Laboratory of Low Dimensional Quantum Physics, Department of Physics, Tsinghua University, Beijing 100084, China}
\affil[3]{Beijing Key Laboratory of Fault-Tolerant Quantum Computing, Beijing 100193, China}
\affil[4]{Institute of Computing Technology, Chinese Academy of Sciences, Beijing 100190, China}
\affil[5]{School of Medical Technology, Beijing Institute of Technology, Beijing 100081, China}
\affil[6]{Institute of Physics, Chinese Academy of Sciences, Beijing 100190, China}
\affil[7]{School of Physical Sciences, University of Chinese Academy of Sciences, Beijing 100049, China}
\affil[8]{CAS Center for Excellence in Topological Quantum Computation, UCAS, Beijing 100190, China}
\affil[9]{Hefei National Laboratory, Hefei 230088, China}
\affil[10]{Frontier Science Center for Quantum Information, Beijing 100184, China}
\affil[*]{Address correspondence to: xuhz@baqis.ac.cn (H.-Z.X.); humj@baqis.ac.cn (M.-J.H.); dongeliu@mail.tsinghua.edu.cn (D.E.L.)}

\date{\today}

\begin{document}

\maketitle

\begin{abstract}
As quantum computing systems continue to scale up and become more clustered, efficiently compiling user quantum programs into high-fidelity executable sequences on real hardware remains a key challenge for current quantum compilation systems. In this study, we introduce a system-software framework that integrates resource virtualization and hardware‑aware compilation for real quantum computing processors, termed \textsc{QSteed}. QSteed virtualizes quantum processors through a four-layer abstraction hierarchy comprising the Real Quantum Processing Unit (QPU), Standard QPU (StdQPU), Substructure of the QPU (SubQPU), and Virtual QPU (VQPU). These abstractions, together with calibration data, device topology, and noise descriptors, are maintained in a dedicated database to enable unified and fine-grained management across superconducting quantum platforms. At run-time, the modular compiler queries the database to match each incoming circuit with the most suitable VQPU, after which it confines layout, routing, gate resynthesis, and noise-adaptive optimizations to that virtual subregion. The complete stack has been deployed on the Quafu superconducting cluster, where experimental runs confirm the correctness of the virtualization model and the efficacy of the compiler without requiring modifications to user code. By integrating resource virtualization with a select-then-compile workflow, \textsc{QSteed} demonstrates a robust architecture for compiling programs on noisy superconducting processors. This architectural approach offers a promising path towards efficient compilation needs across various superconducting quantum computing platforms in the noisy intermediate-scale quantum (NISQ) era.
\end{abstract}


\section{Introduction}
In recent years, various quantum computing platforms, including superconducting qubit~\cite{google2019,ustczcz1,zhangdigital2022,google2024}, ion-trap~\cite{wrightbenchmarking2019,PRXQuantum.2.020343,houindividually2024}, neutral atoms~\cite{bluvsteinquantum2022,everedhigh-fidelity2023,bluvsteinlogical2024}, and photonic quantum devices~\cite{jiuzhang1,jiuzhang3}, have achieved remarkable progress, with increasing qubit control fidelities and scalable architectures reaching dozens or even hundreds of qubits. Despite these advances, quantum computing devices remain scarce resources. The advent of quantum cloud platforms has alleviated this limitation by providing remote access to both public and private quantum hardware via cloud services. This paradigm shift significantly lowers the threshold for quantum experimentation and accelerates progress in domains such as quantum chemistry~\cite{cao2019quantum,doi:10.34133/2020/1486935}, machine learning~\cite{biamontequantum2017,doi:10.34133/research.0134,PERALGARCIA2024100619}, combinatorial optimization~\cite{BLEKOS20241}, and fundamental quantum physics~\cite{fausewehquantum2024,PRXQuantum.5.037001}.
However, deploying high-level quantum algorithms on real hardware remains challenging due to the mismatch between algorithmic abstractions and hardware operations. Bridging this gap depends on efficient quantum compilation, which serves as a critical layer for transforming quantum programs into hardware-executable instructions.

In the field of quantum compilation, several commercial-grade software frameworks have been developed, including IBM’s Qiskit~\cite{Qiskit}, Google’s Cirq~\cite{cirqdevelopers202411398048}, Rigetti’s PyQuil~\cite{smith2016practical}, Quantinuum’s Pytket~\cite{Sivarajah2021}, and Origin Quantum’s QPanda~\cite{dou2022qpanda}. These frameworks provide comprehensive toolchains that are tightly integrated with their respective hardware platforms. Concurrently, academic research has produced a number of specialized toolkits, such as Quartz~\cite{Quartz} for circuit super-optimization and BQSKit~\cite{doecode_58510} which incorporates advanced synthesis algorithms. Additionally, there are compilers tailored to specific quantum algorithms, such as Qcover~\cite{xu2024quafu} for the quantum approximate optimization algorithm (QAOA)~\cite{farhi2014quantum}. Despite these advances, most existing frameworks still face challenges in handling hardware noise and adapting to multi-backend cloud platforms. They generally follow a common paradigm: mapping logical circuits directly onto the entire physical device. However, in the NISQ era, hardware noise on quantum processors (such as two-qubit gate error rates) is non-uniformly distributed. Simply compiling and optimizing circuits by mapping them to the full chip neglects the intrinsic differences of hardware resources. In this context, developing compilation strategies that proactively avoid high-noisy hardware resources emerges as an effective pathway to improving quantum program performance. Furthermore, deploying compilation software on multi-backend heterogeneous quantum cloud platforms and achieving unified resource management and hardware-aware compilation still requires further architectural innovation.

In this work, rather than concentrating on specific compilation optimization algorithms, we focus on an architectural-level innovation in compilation design. We introduce QSteed, a novel quantum compilation system designed for deployment on superconducting quantum computing platforms. QSteed embodies a distinct architectural paradigm built on resource virtualization and a select-then-compile workflow. In particular, it tightly integrates a modular compiler with a quantum resource virtualization layer, featuring several key characteristics: (1) The resource virtualization manager uses heuristic strategies to identify high-quality subregions of a quantum chip, which are then abstracted into a queryable database of virtual quantum processing units (VQPUs). This enables efficient and unified management of multiple quantum backends. (2) The quantum compiler first selects an optimal VQPU for a given input circuit based on structural similarity or fidelity metrics. It then performs efficient, hardware-aware transpilation on this smaller subregion to generate a high-fidelity executable circuit. (3) The system is packaged into two lightweight APIs, one for adding or updating quantum processors in the database and another for user task compilation, simplifying integration with existing superconducting quantum cloud toolchains.

We have deployed QSteed on a quantum cloud computing cluster based on superconducting qubit processors. Specifically, it has been integrated into the Quafu quantum cloud platform, which validates the effectiveness of its overall architecture. To assess QSteed's compilation performance, we executed benchmark circuits with up to 30 qubits on Quafu’s \textit{Baihua} processor and simulated larger circuits using calibration parameters from the Baihua chips. To further examine the robustness of the QSteed architecture, we performed the same simulations with parameter data from Google’s \textit{Willow} processor. Experiments on real hardware and classical simulations show that, for the evaluated small- to medium-scale benchmark circuits, QSteed consistently outperforms leading toolchains such as Qiskit and Pytket in compilation speed while achieving comparable or higher circuit fidelity. Overall, QSteed provides efficient support for practical quantum cloud services. Conceptually, the architecture has the potential to be extended to other quantum computing platforms, but realizing such portability would require redesigning the VQPUs database generation and qubit mapping strategies for each specific hardware architecture. At present, all implementations and experiments are based on superconducting qubit platforms.

\begin{figure*}[ht]
\centering
\includegraphics[width=1\textwidth]{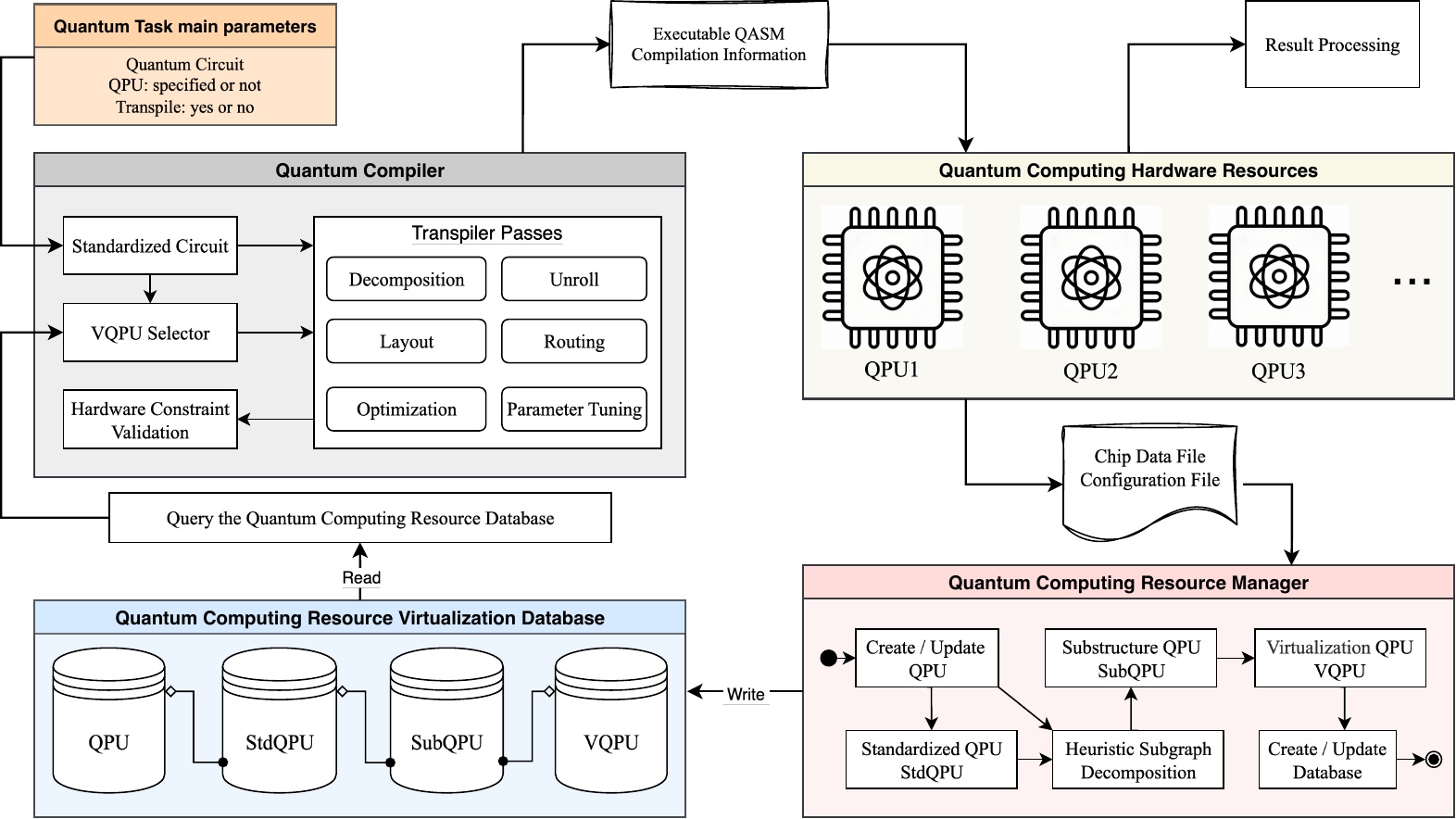}
\caption{QSteed system architecture. It consists primarily of two components: the quantum compiler and the quantum computing resource manager. The manager models the quantum chip at various abstraction levels, including QPU, StdQPU, SubQPU, and VQPU. These representations are stored in a quantum computing resource virtualization database, enabling unified management of quantum backend devices. The compiler queries the virtualization database to compile the user task onto the optimal physical qubits, returning the optimized executable QASM circuit along with relevant compilation information.}
\label{fig-fig1}
\end{figure*}

\section{Results}
\subsection{Overall system architecture design}
As schematically depicted in Figure~\ref{fig-fig1}, the QSteed system is designed with a layered architecture. Its core consists of two key functional modules: a quantum compiler and a quantum computing resource manager. Through a unified API interface, the system enables efficient and seamless hardware deployment. The operational workflow of the system is as follows:

On the hardware management side, a backend operator registers a quantum processor via the \texttt{Manager API}. This process imports essential metadata, such as device topology, real-time calibration data, and native gate sets, into the resource manager. The manager then proactively virtualizes the physical hardware by constructing a four-layer hierarchical model: QPU, StdQPU, SubQPU, and VQPU. These models are stored in a relational database, establishing a pre-computed and queryable resource pool for the compiler, which enables unified management across multiple superconducting backends.

On the quantum task processing side, a user task (containing information such as the quantum circuit and whether a backend is designated) is forwarded to the compiler through the \texttt{Compiler API}. The compiler adopts a \textit{select-then-compile} workflow: First, the input circuit is standardized by rewriting it into a hardware-agnostic unified representation. Next, the compiler queries the resource database to select an optimal VQPU that best matches the circuit's structural or fidelity requirements. Once the target VQPU has been determined, the compiler executes a hardware-aware transpilation process tailored to this optimal subregion, generates Quantum Assembly Language (QASM) \cite{openqasm2,fu2019eqasm} code ready for the target hardware, and returns a detailed compilation report (e.g., compilation time, circuit depth, and gate counts). Finally, it performs hardware constraint verification on the output to ensure the generated QASM code can run safely and directly on the selected quantum processor.

By deeply integrating the compilation process with resource management, QSteed provides unified orchestration of quantum resources and enables efficient task compilation and execution. Its virtualization and select-then-compile mechanism decouples hardware characteristics from user tasks, allowing quantum circuits to be dynamically mapped onto the most suitable physical qubits. Related use cases and implementation details are provided in Section S1 of the Supplementary Materials.

It should be noted that throughout the system design, our primary focus is on two-qubit gate noise. This emphasis is reflected in several core components, including the fidelity-first strategy in resource virtualization, the fidelity-first policy in VQPU selection, and the noise-aware routing algorithm. While decoherence noise is a consideration, it is only incorporated within the routing algorithm through the introduction of gate parallelism considerations.

\subsection{Design of the Quantum Resource Virtualization Manager}
\subsubsection{Hierarchical Abstraction of Quantum Processors}
Our approach to quantum hardware management is conceptually inspired by classical virtualization. In classical computing, virtualization technologies such as virtual machines build an abstraction layer on top of reliable and homogeneous physical resources, thereby enabling efficient resource sharing and isolation. Similarly, our framework virtualizes quantum processors to mask the underlying physical complexity, manage resources more effectively, and provide a unified backend interface for upper-layer compilation services.

However, the analogy ends at this high-level concept, as the motivations and challenges of quantum virtualization are fundamentally different. Classical virtualization is designed to share stable and homogeneous resources, whereas in the NISQ era quantum virtualization must contend with noisy and heterogeneous resources. The central challenge is not simply partitioning a high-quality whole, but identifying and exploiting relatively high-quality subregions within a chip that is inherently imperfect and subject to time-varying noise. Unlike classical virtualization, which seeks to emulate a complete and stable runtime environment, our quantum virtualization constructs a lightweight hardware abstraction layer that is noise-aware and topology-constrained, specifically tailored to support resource management and compilation optimization. This distinction underpins the design of our framework.

As quantum chips scale up and their architectures become increasingly complex, managing quantum resources across multi-backend clusters is becoming progressively more challenging. To address this, we have developed a virtualization manager that abstracts quantum chips into a database representation, enabling efficient orchestration of large-scale quantum clusters. Within this framework, quantum chips are represented through four hierarchical abstraction levels: QPU, StdQPU, SubQPU, and VQPU.

\textbf{QPU.} 
The QPU serves as the foundational abstraction layer, providing a direct digital representation of a physical quantum processor's characteristics. This ``digital twin'' encapsulates critical hardware parameters, including the chip's architecture type, qubit connectivity topology, real-time calibration data (e.g., gate fidelities and coherence times), the supported native gate set, gate durations, a unique hardware identifier, and its current operational status. This layer grounds the entire virtualization stack in the precise state of the physical device.

\textbf{StdQPU.} 
Due to current limitations in micro-/nanofabrication processes, superconducting quantum chips inevitably have defects~\cite{krantz2019quantum}, rendering some qubits unusable and resulting in irregular chip structures. We therefore introduce StdQPU, an idealized standard QPU architecture designed to embed these defective chips. For instance, on superconducting platforms, StdQPU can be modeled as a two-dimensional grid structure, into which most superconducting chips can be mapped. This abstraction facilitates the chip to be dynamically partitioned into multiple regions capable of hosting concurrent workloads, thereby laying the groundwork for QSteed’s future support of multi-program quantum execution~\cite{Niu2023enablingmulti}. Although the present study does not explore such multi-program scenarios in depth, we retain this layer of abstraction to preserve the completeness of the virtualization stack.

\textbf{SubQPU.} 
SubQPU represents high-quality substructures identified within a QPU. For a topology graph $G(V, E)$ with $N$ qubits and $M$ coupling edges, enumerating all its substructures has an exponential complexity of $O(2^{N+M})$. Consequently, we employ heuristic algorithms to identify valuable SubQPUs. To maximize the diversity of these substructures to suit different types of quantum circuits, we designed three complementary heuristic strategies:

\textit{Fidelity-first strategy.} 
This strategy aims to identify $n$-qubit coupled substructures with high average fidelity, which is crucial for noise-sensitive quantum applications. Specifically, for $n=1$ and $n=2$, we construct the optimal substructures by sorting nodes or edges based on their respective single- or two-qubit gate fidelities. For $n \geq 3$, we heuristically construct an $n$-qubit substructure, $subG$, by initializing it with the highest-fidelity edge and iteratively adding adjacent qubits sorted by two-qubit gate fidelity until the substructure reaches size $n$.

\textit{Degree-first strategy.} 
This strategy prioritizes node degree to identify substructures composed of highly connected qubits, which is beneficial for algorithms requiring a high degree of entanglement. Its implementation is similar to the fidelity-first strategy, with the key difference being that during the construction of the substructure $subG$, adjacent qubits sorted by node degree are iteratively added. 

\textit{Random selection strategy.} 
By introducing randomness, this strategy explores substructures with different topologies to enrich the database. In each step of the iteration, a neighboring qubit is randomly added to $subG$.

\begin{figure*}[!ht]
\centering
\begin{minipage}{0.65\textwidth}
\begin{algorithm}[H]
    \caption{Find Substructures in QPU Coupling Graph}\label{alg-subqpu}
    \begin{algorithmic}[1]
        \Require Qubit coupling graph of QPU $G(V, E)$
        \Ensure List of all substructures $all\_substructure$
        \State $N \gets \text{len}(G.nodes())$
        \State $Sn \gets \text{sorted}(G.nodes(), \text{fidelity})$ \Comment{Sort nodes by fidelity}
        \State $Se \gets \text{sorted}(G.edges(), \text{fidelity})$ \Comment{Sort edges by fidelity}
        \State $all\_substructure = [\ ]$
        \State $all\_substructure.\text{append}((1, Sn))$
        \State $all\_substructure.\text{append}((2, Se))$
        \For{$n \gets 3$ to $N$}
            \State $n\_substructure = [\ ]$
            \For{\textbf{each} method in \{\textit{fidelity}, \textit{degree}, \textit{random}\}}
                \For{\textbf{each} edge in $Se$}
                    \State $subG$ $\gets$ Graph()
                    \State $subG.\text{add\_edge}(\textit{edge})$
                    \State $qubits \gets 2$
                    \While{$qubits \neq n$}
                        \State $neighbors \gets \text{sorted}($
                        \Statex \hspace{110pt} $subG.neighbors(), \text{method})$
                        \State $best\_neighbor \gets neighbors[0]$ 
                        \State $subG.\text{add\_node}(best\_neighbor)$
                        \State $qubits \gets qubits + 1$
                    \EndWhile
                    \State $n\_substructure.\text{append}(subG)$
                \EndFor
            \EndFor
            \State $n\_substructure \gets \text{sorted}(n\_substructure, \textit{ave\_fidelity})$
            \State $all\_substructure.\text{append}((n, n\_substructure))$
        \EndFor
    \end{algorithmic}
\end{algorithm}
\end{minipage}
\end{figure*}


\begin{figure*}[ht]
\centering
\includegraphics[width=0.8\textwidth]{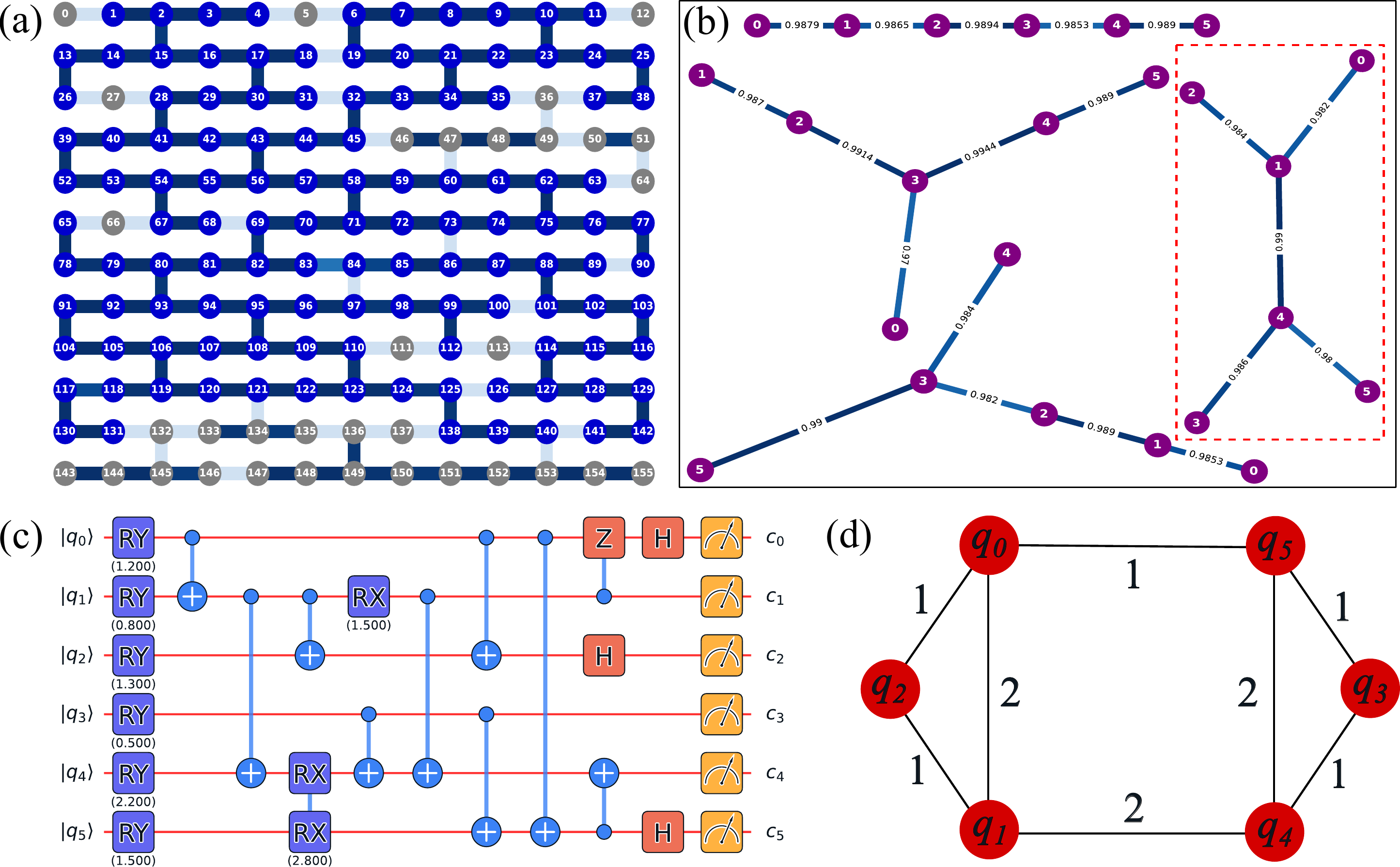}
\caption{(a) The schematic of the Baihua quantum chip, with blue regions representing 122 connected qubits.
(b) The 6-qubit VQPU structure within the Baihua quantum chip, where the red box highlights the target VQPU (the mapping from virtual qubits $v$ to physical qubits $q$ is $v_{\{0,1,2,3,4,5\}} \to q_{\{61,62,63,74,75,76\}}$) with the structure most similar to the circuit in (c). (c) Quantum circuit. (d) Weighted graph representation of the quantum circuit, where the nodes represent qubits, the edges represent two-qubit gates, and the edge weights represent the number of two-qubit gates.}
\label{fig-dag-graph}
\end{figure*}

\textbf{VQPU.} 
To unify the description of quantum resources, each SubQPU is further abstracted into a VQPU, the final compiler‐facing representation. The VQPU omits certain physical details of the quantum chip, retaining only the essential information needed for compilation. This includes the mapping from virtual qubits (numbered starting from 0) to physical qubits, the noisy topological structure, the native gate set, and the identifier of the parent QPU. This abstraction supplies the compiler with a clean, standardised interface. Figure~\ref{fig-dag-graph}b shows 6-qubit VQPU configuration extracted from the Baihua chip.

\subsubsection{Complexity Analysis of Virtualization Strategy}
The time complexity of quantum hardware virtualization is primarily composed of two components. The first corresponds to identifying the optimal substructures of the chip, as described in Algorithm~\ref{alg-subqpu}. The second pertains to the deletion and insertion costs incurred during database updates.

For a quantum processor with $N$ qubits and $M$ coupling edges, the time complexity of Algorithm~\ref{alg-subqpu} is dominated by the main loop structure. After an initial sorting of nodes and edges, which takes $O(N \log N + M \log M)$, the algorithm iterates through each of the $M$ edges to construct substructures of size $n$ (from 3 to $N$). The core of this process is an iterative growth step, where new qubits are added to the substructure. By employing a priority queue to efficiently select the best neighboring qubit at each step, the inner loop has a complexity of approximately $O(n \log N)$. Consequently, the total time complexity of the algorithm is approximately $O(M N^2 \log N)$. For sparsely coupled superconducting chips, we have $M = O(N)$, and the complexity simplifies to $O(N^{3}\log N)$. A more detailed derivation can be found in Section S2.1 of the Supplementary Materials. In addition, we performed numerical simulations on representative superconducting chip topologies, including the square-lattice topology adopted by Google and USTC’s Zuchongzhi processor, the heavy-hexagonal topology promoted by IBM, as well as the hexagonal topology. The numerical results (see Figure S2-S5 in the Supplementary Materials) show that for chips with up to 200 qubits, our algorithm can complete the identification of optimal subregions within a few minutes, and the fitted scaling is consistent with $O(N^3 \log N)$. This demonstrates that the approach is fully practical for current NISQ devices.

The update time of the database depends on the specific implementation and architecture adopted. In our current implementation, a MySQL database is used, where record deletion and insertion are carried out in a serial manner. Numerical results (see Figure S3-S5 in the Supplementary Materials) indicate that the associated complexity scales as $O(N^4)$. Possible directions for future improvement may include adopting primary-key–based differential deletion and batch insertion strategies, as well as redesigning the database architecture or considering migration to more efficient systems such as MongoDB or Redis.


\subsection{Design of a Modular Hardware-Aware Quantum Compiler}
\subsubsection{Standardized Circuit}
Before executing the core matching and compilation workflow, QSteed first normalizes user-submitted quantum circuits via a standardized preprocessing module. By correcting common formatting issues, such as invalid measurements, missing classical registers, and redundant qubits, this module effectively ensures that subsequent processes execute reliably. Its primary functions include:

\textit{Measurement and classical-register correction.} 
Quantum programs require classical registers to store measurement results. This module automatically analyzes the QASM code, appends any omitted measurement instructions, and provisions the requisite classical registers, thereby averting runtime errors attributable to user oversight.

\textit{Redundant qubit cleaning.} 
By analyzing the QASM code, this module identifies and removes qubits that do not participate in any effective quantum operations within the circuit. This step reduces unnecessary computational overhead in later compilation phases, conserves scarce qubit resources on NISQ devices, mitigates potential qubit interference, and ultimately helps improve circuit fidelity.

\subsubsection{VQPU Selector}
Benefiting from the construction of the VQPU database, we no longer need to consider the low-level particulars of the entire quantum chips. Instead, we can focus on matching a suitable VQPU for each quantum circuit. The VQPU-selection module begins by filtering all candidate VQPUs from the database that meet the circuit's qubit count requirement. Subsequently, one of the following strategies is employed to determine the final VQPU:

\textit{Fidelity-first strategy.} 
This strategy selects the VQPU with the highest overall fidelity, defined as the product of all two-qubit gate fidelities within the VQPU. Since the VQPU database is pre-sorted by this metric during its construction, this strategy ensures rapid identification of the optimal VQPU for the task.

\textit{Structure-first strategy.} 
This strategy aims to reduce the overhead of SWAP gates generated during the subsequent qubit mapping process by identifying the  VQPU that most closely resembles the circuit topology. This is a subgraph isomorphism problem within the field of quantum compilation~\cite{li2020qubit,park2022fast,ge2024quantum}. In QSteed, we employ the following practical implementation: 
We represent both the quantum circuit and the VQPU as weighted graphs. In the circuit graph, vertices correspond to logical qubits, and edges correspond to two-qubit gates (multi-qubit gates are expanded into a fully connected subgraph), where edge weights signify the number of two-qubit gates. For example, the quantum circuit in Figure~\ref{fig-dag-graph}c can be represented as the weighted graph shown in Figure~\ref{fig-dag-graph}d. Similarly, in the VQPU graph, vertices denote physical qubits, edges represent direct coupling pairs, and edge weights indicate the fidelity of two-qubit gates. For structural comparison, we first normalize the edge weights in both graphs to the range [0,1]. Then, we query whether there exists a graph isomorphic to the circuit diagram among the candidate VQPUs (ignoring edge weights). If such an isomorphic match is found, that VQPU is selected, enabling perfect mapping of the quantum circuit onto the hardware without requiring additional SWAP gates. When no isomorphic match exists, we quantify graph similarity with the Weisfeiler–Lehman subtree kernel~\cite{wlKernel} and select the VQPU that yields the highest similarity score. In cases where multiple VQPUs have the same score, we break ties by choosing the VQPU with the highest overall fidelity. Detailed algorithmic descriptions appear in Section S2 of the Supplementary Materials.

\subsubsection{Quantum Circuit Transpiler}
Once the target VQPU has been identified, the quantum circuit transpiler is responsible for compiling the logical circuit into a physical circuit that can be executed on that VQPU. The key challenge of this process is to satisfy the hardware's topological constraints while performing gate-level optimizations to enhance the fidelity of the final circuit. To facilitate flexible and efficient compilation, the transpiler of QSteed adopts a modular pipeline architecture and employs directed acyclic graph (DAG) as the intermediate representation of circuits, which aligns with the design philosophy of mainstream frameworks such as Qiskit~\cite{Qiskit} and Pytket~\cite{Sivarajah2021}. This design offers a high degree of flexibility and extensibility, enabling researchers to conveniently test new compilation algorithms.

\begin{figure}[ht]
\centering
\includegraphics[width=0.48\textwidth]{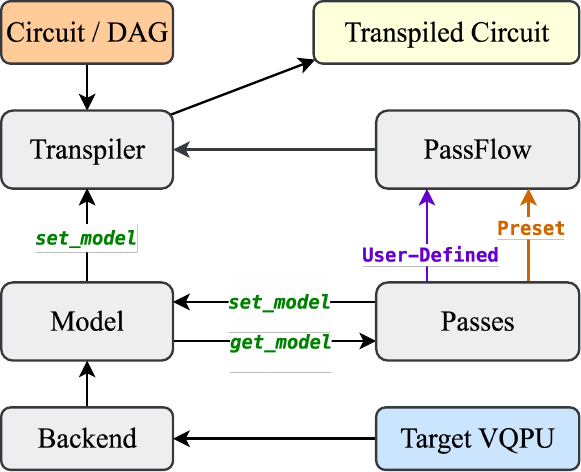}
\caption{The quantum circuit transpilation and optimization workflow adopts a modular design, consisting of core components: \texttt{Transpiler}, \texttt{Pass}, \texttt{PassFlow}, \texttt{Model}, and \texttt{Backend}. The target VQPU is first modeled as a \texttt{Backend} object and then abstracted into a \texttt{Model} object. During the transpilation process, the \texttt{Model} object stores and passes the parameter data required for transpilation. The transpiler selects a predefined or user-defined \texttt{PassFlow} based on the optimization level to execute the transpilation process. A \texttt{PassFlow} is composed of a sequence of functional components, each responsible for a specific transpilation task. The transpilation process is primarily performed on the directed acyclic graph (DAG) representation of the quantum circuit. Upon completion, the transpiled quantum circuit is output.
}
\label{fig-transpiler}
\end{figure}

Our transpiler is specifically composed of several key components: Transpiler, Pass, PassFlow, Model, and Backend, as depicted in Figure~\ref{fig-transpiler}. The \texttt{Transpiler} serves as the main entry point, responsible for selecting a particular \texttt{PassFlow} to execute the transpilation process according to a preset optimization level. Users also have the option to customize their own \texttt{PassFlow}. A \texttt{PassFlow} is an ordered sequence of various \texttt{Pass} instances, each responsible for a specific task such as gate decomposition, qubit mapping, circuit optimization, and variational circuit parameter tuning. Users can create a custom \texttt{Pass} by inheriting from the \texttt{BasePass} class to implement specific functionalities. Throughout transpilation, a shared \texttt{Model} object maintains all state information, including the current qubit layout and the \texttt{Backend} description derived from the selected VQPU. Each \texttt{Pass} obtains this state via \textit{get\_model}, performs its transformation, and writes back the updated state using \textit{set\_model}. This disciplined data flow keeps the \texttt{Model} synchronized across the pipeline, ensuring that every subsequent pass operates on the latest information.

Among all transpilation steps, qubit mapping is often the most critical, directly determining the final circuit's depth and fidelity. This process dynamically assigns logical qubits to physical qubits and is typically divided into two key stages, layout and routing. Layout establishes the initial mapping of logical qubits to physical qubits, while routing ensures that all two-qubit gate operations adhere to the hardware's connectivity constraints by inserting SWAP gates. 
Accordingly, the remainder of this subsection focuses on the qubit-mapping process; the other transpilation steps follow standardized methodologies detailed in Section S3 of the Supplementary Materials.  

Numerous qubit mapping algorithms have been developed~\cite{kusyk2021survey,zhu2025quantum}, among which the SABRE algorithm~\cite{10.1145/3297858.3304023} is widely adopted in mainstream compilation frameworks like Qiskit due to its recognized efficiency and scalability. However, this algorithm still has certain limitations, such as its trivial initial layout and its insensitivity to hardware noise. Currently, various optimization algorithms specifically targeting initial layout~\cite{zhu2020dynamic,park2022fast,huang2022reinforcement}, as well as routing algorithms that consider hardware noise~\cite{murali2019noise,niu2020hardware,escofet2024route}, have been proposed. The qubit mapping algorithm implemented in QSteed introduces the following improvements over the original SABRE algorithm:

\textit{Structure-aware initial layout.} 
Let the quantum circuit’s weighted graph be $G_{\text{QC}}$ and the graph of the optimal VQPU determined by the VQPU selector be $G_{\text{VQPU}}$. We apply two layout initialization methods: (1) Degree initialization: Nodes in $G_{\text{QC}}$ and $G_{\text{VQPU}}$ are sorted by degree, and logical qubits $q$ are mapped one-to-one onto virtual qubits $v$ based on the sorted order. For instance, if the sorted orders are $[q_1, q_3, q_0, q_2]$ for $G_{\text{QC}}$ and $[v_2, v_0, v_3, v_1]$ for $G_{\text{VQPU}}$, the initial layout is $q_{\{1,3,0,2\}} \leftrightarrow v_{\{2,0,3,1\}}$. (2) Weight initialization: If node degrees are identical, sorting is performed based on edge weights, and nodes are then mapped one-to-one.

\textit{Noise-aware routing.} 
The original SABRE algorithm employs the nearest neighbor cost function, defined by the distance matrix $D[i][j]$, which represents the shortest path between physical qubits $Q_i$ and $Q_j$. However, for NISQ hardware with non-uniform two-qubit gate fidelities, this approach may fail to maximize the overall circuit fidelity. To improve this, we replace the distance matrix $D[i][j]$ with a fidelity matrix $\text{Fi}[i][j]$, where each entry identifies the path with the highest cumulative fidelity, calculated as the product of all two-qubit gate fidelities along the path. 
We then define an improved, noise-aware heuristic  cost function $H_\text{Fi}$,
\begin{equation}
\begin{array}{r}
H_\text{Fi} = max \left(\operatorname{decay}\left(\text{SWAP}.q_1\right), \text{decay}\left(\text{SWAP}.q_2\right)\right) \\
*\left\{\frac{1}{|F|} \sum_{\text {gate} \in F} \text{Fi}\left[\pi\left(\text{gate.}q_1\right)\right]\left[\pi\left(\text{gate.}q_2\right)\right]\right. \\
\left.+W * \frac{1}{|E|} \sum_{\text{gate} \in E} \text{Fi}\left[\pi\left(\text{gate.}q_1\right)\right]\left[\pi\left(\text{gate.}q_2\right)\right]\right\},
\end{array}
\label{eq-HFi}
\end{equation}
where $F$ represents the front layer of gates in the quantum circuit, $E$ denotes the extended layer of gates, and $\pi()$ represents the mapping from logical qubits $q_{\{1,2,\ldots,n\}}$ to physical qubits $Q_{\{1,2,\ldots,N\}}$. The parameter $W \in (0,1)$ is a weight factor, and $\text{decay}(q_i) = 1 + \delta$ characterizes the parallelism of the compiled circuit. Detailed explanations of these parameters can be found in \cite{10.1145/3297858.3304023}. 

However, solely pursuing the highest-fidelity path may lead to an increased number of SWAP operations, which in turn deepens the circuit and exacerbates decoherence. Therefore, we further design a hybrid heuristic $H_{\text{M}}$ that integrates both distance-based and fidelity-based cost functions. Specifically, at each step of the routing process, given the set of candidate SWAPs $S$, we first compute the distance-based cost $H_{\text{D}}$ (replacing $\text{Fi}[i][j]$ in Eq. $\ref{eq-HFi}$ with the distance matrix $\text{D}[i][j]$) for each SWAP and retain only those with the minimum value. If multiple candidates remain, we evaluate the fidelity-based cost $H_{\text{Fi}}$ within this subset and select the SWAP with the highest score. If a tie still exists, one SWAP is selected at random from the remaining candidates as the final result, denoted by $s_{\text{opt}}$. This method is formally defined as:
\begin{equation}
s_{\text{opt}} := \operatorname{Random} \left( 
\operatorname*{arg\,max}_{s \in \operatorname*{arg\,min}_{s \in S} H_{\text{D}}(s)} H_{\text{Fi}}(s) 
\right).
\end{equation}

\subsubsection{Hardware Constraint Validation}
Before dispatching the compiled QASM code to the target backend, QSteed performs a series of final hardware compatibility checks. This step does not constitute formal program verification in the strict academic sense~\cite{liu_formal_2019,10.1145/3453483.3454061,doi:10.1142/S0219749908003530}; rather, it is a pragmatic pre-execution safeguard commonly embedded in production-grade compilation frameworks. Its purpose is to intercept tasks that are destined to fail because they violate the physical limitations of the hardware, thereby averting futile consumption of real quantum resources. Specifically, QSteed's hardware constraint validation module includes:

\textit{Two-qubit coupling constraints.} 
Verifies that every two-qubit gate in the circuit respects the qubit-coupling graph of the target processor, thereby ensuring topological compatibility with the designated device.

\textit{Gate count and circuit depth limits.} Checks that the total number of gates and the circuit depth remain below the backend’s execution thresholds, preventing invalid tasks with excessively low fidelity.

\textit{Non-empty circuit check.} 
Confirms that the circuit contains at least one quantum operation, averting the submission of empty jobs that would waste quantum resources and trigger runtime errors.

\begin{figure*}
\centering
\includegraphics[width=1\textwidth]{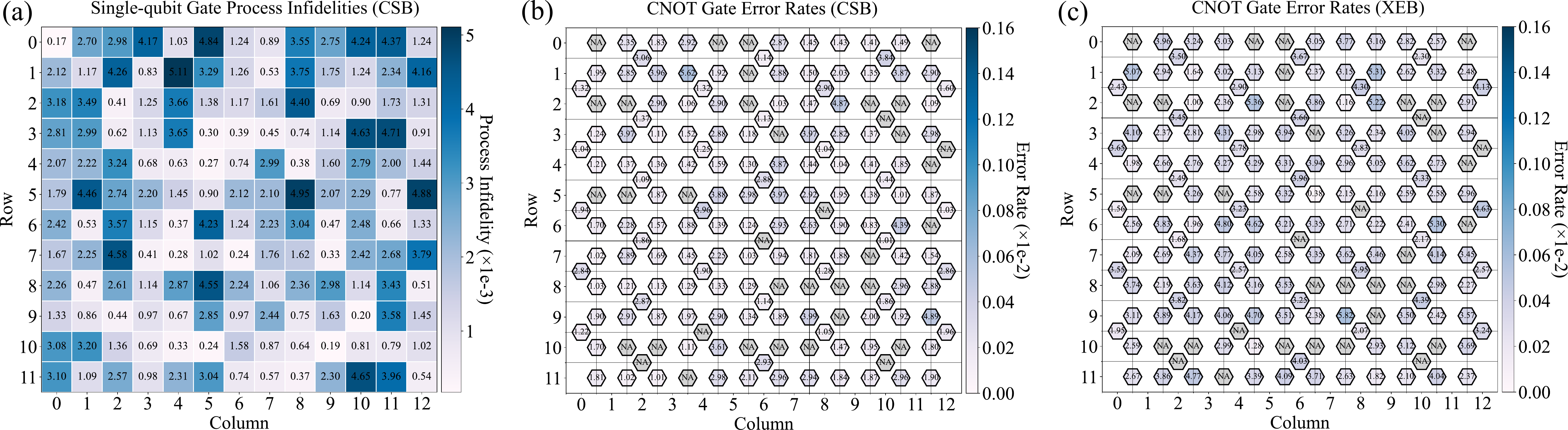}
\caption{Benchmarking results of the Baihua quantum chip. (a) Single-qubit process infidelities from CSB. (b) CNOT gate error rates from CSB. (c) CNOT gate error rates from XEB. NA means the error rate is too high or gates are unavailable.}
\label{fig-bm}
\end{figure*}

\subsection{Experiments and Simulations}
We first conducted real-device experiments using the Quafu superconducting quantum computing cloud platform, selecting the Baihua quantum processor (as shown in Figure~\ref{fig-dag-graph}a) as the target backend. Through the cloud platform, we accessed the most recent calibration data of the chip, including qubit coupling topology as well as single- and two-qubit gate fidelities. The median values of key performance parameters for the Baihua chip are a single-qubit error rate of $7.6\times 10^{-4}$, a two-qubit error rate of $2\times 10^{-2}$, a relaxation time $T_1$ of $71.608~\mu s$, and a dephasing time $T_2$ of $23.724~\mu s$ (these values may vary after recalibration of the chip). To ensure the reliability of the calibration data provided by the platform, we employed the ErrorGnoMark~\cite{ErrorGnoMark_new} software for benchmarking, which is designed to provide a comprehensive diagnostic of quantum chips.
Specifically, we assessed single-qubit gate and CNOT gate infidelities using cross-entropy benchmarking (XEB)~\cite{Arute2019} and channel spectrum benchmarking (CSB)~\cite{Gu2023}, thereby establishing a reliable hardware-noise baseline for the subsequent experiments. Although CSB can characterize process infidelity, statistical infidelity, and rotation-angle errors, our compilation study concerns only gate fidelity, so we report process infidelity exclusively. Gate benchmarking results are summarized in Figure~\ref{fig-bm}. Specifically, Figure~\ref{fig-bm}a illustrates single-qubit process infidelities obtained through CSB, which approximate average gate infidelities. CNOT process infidelities, also derived from CSB, are reported in Figure~\ref{fig-bm}b. Figure~\ref{fig-bm}c details average CNOT error rates from XEB, reflecting the difference between measured and ideal output distributions. A theoretical expectation of consistency exists for the data presented in Figure~\ref{fig-bm}b and c. The observed discrepancies may be attributable to noise instability.

\begin{figure*}[ht]
\centering
\includegraphics[width=1\textwidth]{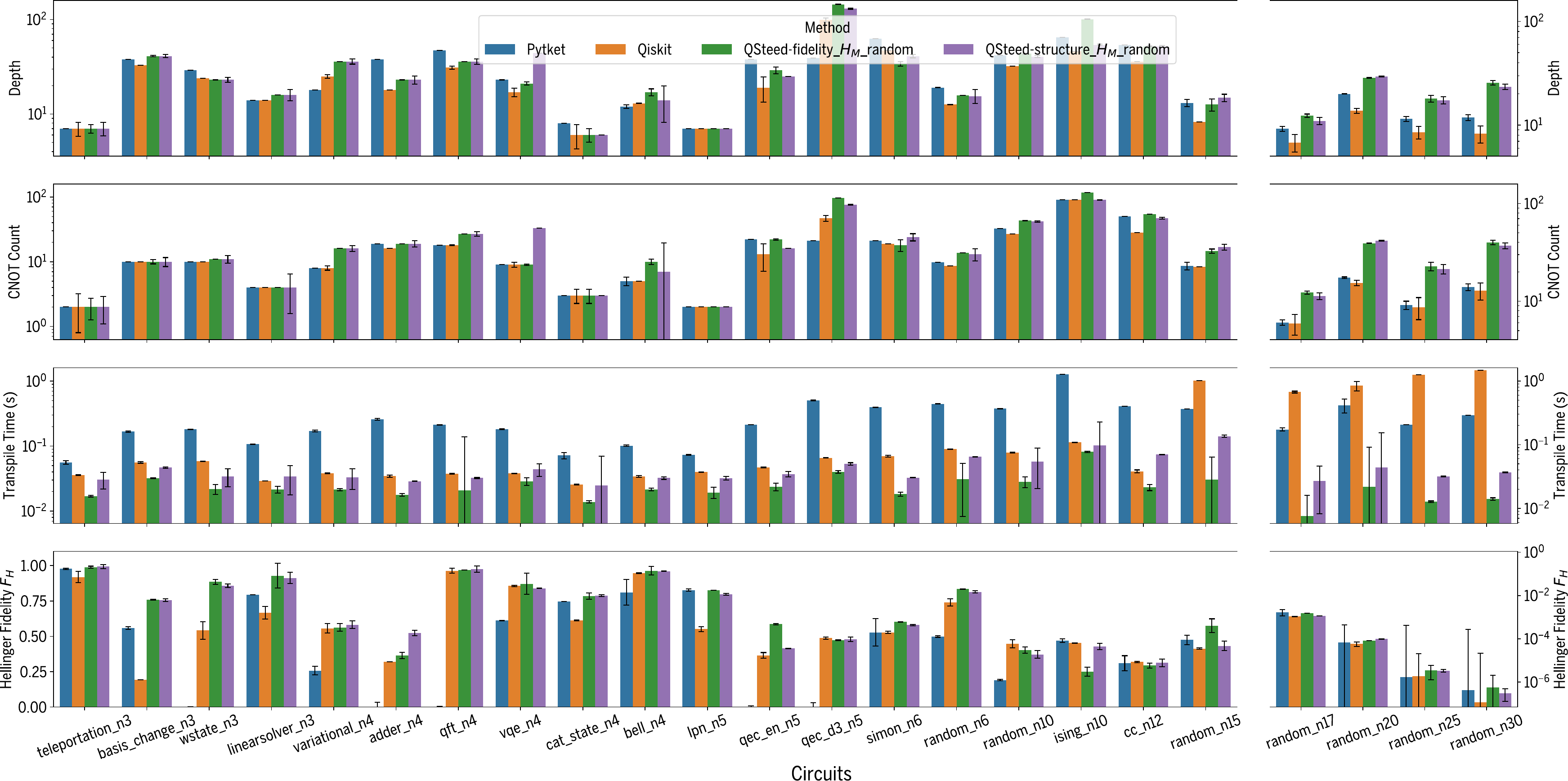}
\caption{Performance comparison of different quantum compilers on the Baihua quantum processor. The results for each benchmark circuit represent the median of 5 runs, while the results for random types are averaged over more than 50 runs of randomly generated circuits. Error bars indicate the standard error of the mean. The vertical axes, from top to bottom, represent the compiled circuit depth, the number of CNOT gates, the transpilation time, and the Hellinger fidelity $F_H$ (higher is better). Except for the random circuits, the circuit names (trailing number denotes qubits) on the horizontal axis are derived from the QASMBench benchmarking suite~\cite{10.1145/3550488}.
}
\label{fig-small-baihua}
\end{figure*}

To ensure fairness in performance comparisons among different compilers under realistic noise conditions, all frameworks were evaluated using the same calibration data obtained from the Baihua chip. Subsequently, all benchmarking runs for QSteed, Qiskit, and Pytket were executed within a single continuous time block, ensuring that they operated under the same physical hardware state and thereby minimizing the impact of calibration drift. In addition, to avoid the influence of short-term chip fluctuations, each data point for algorithm-specific circuits is reported as the median of five consecutive runs, while results for random circuits are averaged over approximately 50 runs, providing statistically robust performance metrics. The compilation results are presented in Figure~\ref{fig-small-baihua}.
Across most evaluated benchmarks, QSteed achieves shorter compilation times while maintaining comparable or higher Hellinger fidelity. This advantage stems from its pre-built VQPU database, which restricts optimization to an appropriate VQPU rather than the entire chip. In contrast, Qiskit and Pytket take the entire chip as input, which prolongs their compilation time. Moreover, although QSteed does not lead to shallower circuits or fewer gates, its noise- and topology-aware VQPU selection strategy, coupled with a noise-aware qubit mapping pass, systematically yields higher Hellinger fidelity than either Qiskit or Pytket, with most of the fidelity gains attributable to the former. Notably, the fidelity-first strategy compiles more quickly because it simply selects the VQPU with the highest aggregate fidelity, whereas the structure-first strategy incurs additional overhead to compute graph-similarity scores in order to identify the best-matching VQPU. However, when a circuit is exactly isomorphic to a candidate VQPU, such as the linear \texttt{ising-n10} circuit, the structure-first strategy achieves higher Hellinger fidelity. Detailed results for QSteed under other compilation strategies are provided in Figure S6 of the Supplementary Material.

\begin{figure*}[ht]
\centering
\includegraphics[width=1\textwidth]{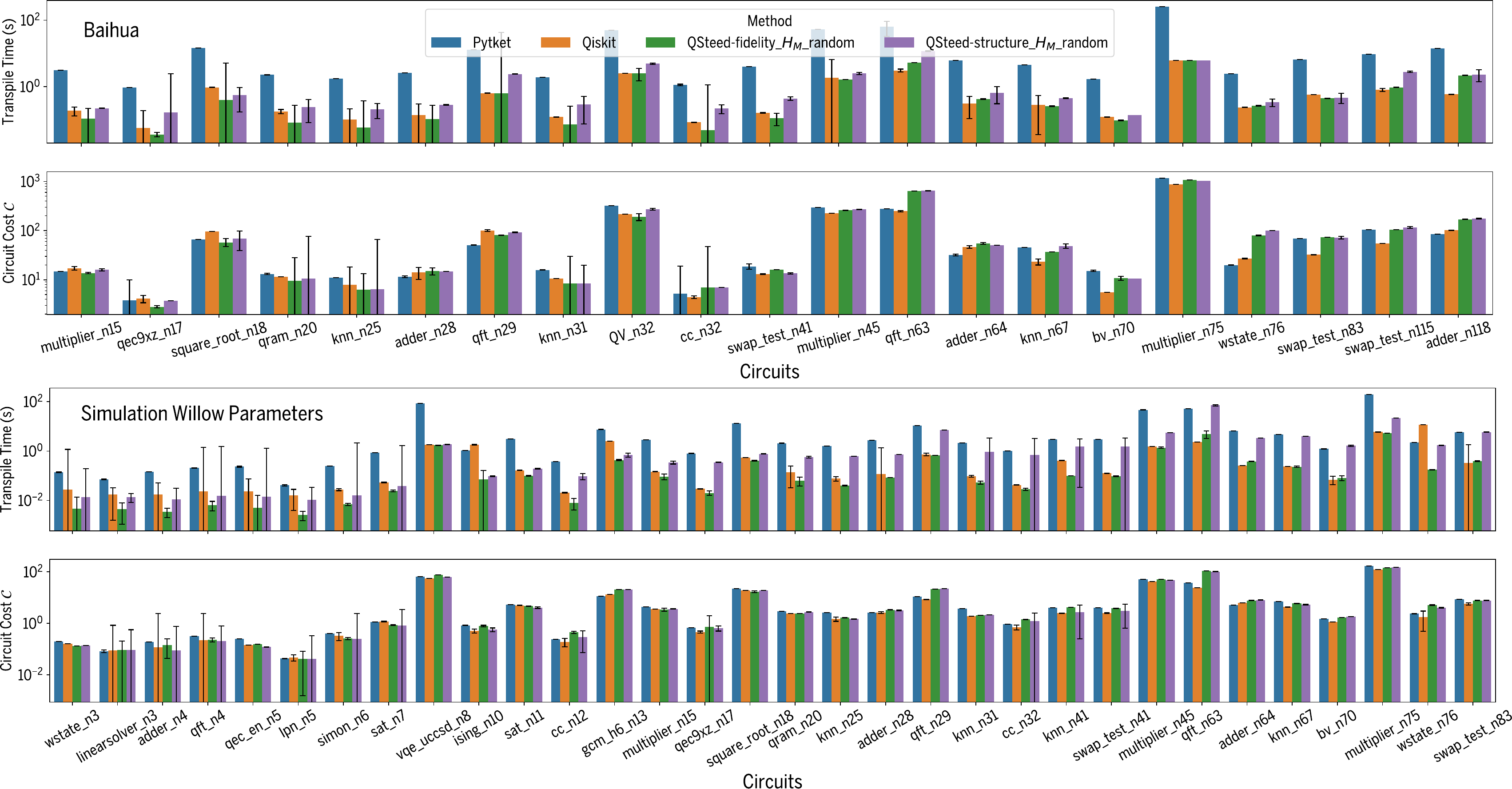}
\caption{Simulation results. For large-scale quantum circuits, the sampling results obtained from NISQ devices are unreliable, and obtaining ideal results from simulators is also challenging, which hinders the calculation of the Hellinger fidelity $F_H$. Therefore, we numerically simulate the circuit cost function $\mathcal{C}$ for each compiled circuit based on the qubit fidelity information of the Baihua quantum processor. To further evaluate the robustness of our compilation framework, we also compute $\mathcal{C}$ based on the Willow quantum processor parameters reported in ~\cite{google2024}. The upper and lower panels respectively present the performance comparison of various quantum compilers on the Baihua processor, and comparison of the simulation results based on data of Willow processor. For each benchmark circuit, results represent the median over 5 runs. Error bars indicate the standard error of the mean. Lower values of transpilation time and the circuit cost function $\mathcal{C}$ indicate better compilation performance. The circuit names (trailing number denotes qubits) along the horizontal axis are taken from the QASMBench benchmarking suite~\cite{10.1145/3550488}.}
\label{fig-baihua-willow}
\end{figure*}

For larger-scale quantum circuits, direct execution on physical hardware becomes unreliable due to the cumulative impact of hardware noise, and simulation for obtaining ideal results presents significant computational demands. Therefore, we use the circuit cost function, $\mathcal{C}$, as a proxy for the expected circuit fidelity to evaluate compilation performance for larger circuits. The upper panel of Figure~\ref{fig-baihua-willow} shows benchmarking results for circuits scaling from a dozen to over a hundred of qubits. Based on characterization data from the Baihua quantum processor, parameters in Eq.~\ref{eq-circuit-cost} were set to $F_i^{1q} = 0.996$ and $K = 0.995$. The analysis demonstrates that for moderate-scale circuits (up to $\approx$ 50 qubits), QSteed employing the fidelity-first strategy offers the shortest compilation times while maintaining circuit cost $\mathcal{C}$ comparable to, or lower than, that of Qiskit and Pytket. However, as circuit size increases, QSteed’s compilation speed advantage over Qiskit gradually wanes. This is primarily because Qiskit has migrated its performance-critical kernels to Rust, wrapped by a Python interface,  whereas QSteed has so far refactored only the DAG and qubit mapping modules in C++. Secondly, particularly when the circuit's qubit count approaches the chip’s full capacity, searching for an optimal mapping over the entire processor becomes essentially as costly as searching within the VQPUs subspace. Moreover, for large-scale circuits, Qiskit generally achieves lower $\mathcal{C}$. This is primarily attributed to its deeper optimization passes, such as those utilizing commutation relations and advanced circuit synthesis, while the current QSteed only employs optimizations like gate elimination and merging.

To validate QSteed’s robustness across diverse processor architectures, we conducted the same simulations on Google’s 105-qubit superconducting Willow processor, which features a 2D grid topology (using the 97-qubit data available in \cite{google2024}).  Employing chip parameters from \cite{google2024}, the constants in Eq.~\ref{eq-circuit-cost} were set to $F_i^{1q}=0.9996$ and $K=0.995$. The lower panel of Figure~\ref{fig-baihua-willow} shows the results. On Willow, we observed similar trends to those on Baihua: QSteed demonstrated an advantage in compilation speed while achieving comparable or superior circuit cost for small to medium-scale circuits. 

These results indicate that QSteed’s virtualization mechanism and select-then-compile workflow are robust. Moreover, the consistent benefits observed on two distinct superconducting processor structures indicate that the framework’s advantages are not confined to a single hardware topology, although validating its applicability across other quantum hardware platforms remains future work.

\section{Discussion}
In this work, we have designed and implemented QSteed, a novel compilation architecture that abstracts complex quantum hardware into a database of high-quality VQPUs. The data associated with these VQPUs is updated upon hardware recalibration. This abstraction obviates the need for the compiler to directly interface with the entire complex hardware backend. Instead, it maps user tasks to the optimal VQPU within the database, thereby enabling an efficient select-then-compile workflow. Experimental and simulation results demonstrate that the QSteed compiler consistently outperforms mainstream compilers such as Qiskit and Pytket on most benchmark circuits, showing strong practicality. 

We also acknowledge that QSteed faces scalability challenges when considering future fault-tolerant quantum computers that may comprise thousands or even millions of qubits. At such scales, even the offline precomputation of the quantum resource virtualization database could become a bottleneck, suggesting that it may be necessary for more scalable or parallelizable heuristics for optimal VQPU identification. Potential directions include restricting the search depth, exploiting regularities in chip topologies for optimization, and parallelizing the search from multiple starting points.

Another limitation concerns calibration drift. Although the VQPU database stays dynamically synchronized with calibration data, if the backend hardware is not recalibrated in a timely manner, our current system architecture cannot detect parameter drift. This is a general issue that cannot be fully resolved at the compilation or virtualization level alone, but requires coordination with hardware benchmarking and automated calibration systems. In future work, we plan to integrate our system architecture with such benchmarking and calibration mechanisms to mitigate this limitation.

In addition, although our implementation and validation have primarily targeted superconducting processors, the architectural paradigm of QSteed holds potential for broader applicability. However, extending QSteed to other quantum hardware platforms requires addressing two key challenges:

First, applying QSteed's resource virtualization mechanism to other hardware platforms necessitates the identification of high-quality subregions within the specific hardware architecture and the corresponding construction of a VQPU database. Below, we explore possible adaptation pathways for several leading platforms: For one-dimensional linear ion-trap, although they are logically fully connected, the fidelity of two-qubit gates varies significantly with ion position~\cite{wrightbenchmarking2019}. Therefore, a fidelity-first strategy could still be envisioned to identify and extract SubQPUs composed of ions linked by high-fidelity connections. For modular quantum charge-coupled device (QCCD) architectures, which consist of multiple traps where ions are fully connected within each trap but have limited connectivity between traps~\cite{pino2021demonstration}, we can foresee that a degree-first strategy would be an effective heuristic for finding the most highly connected physical regions to serve as SubQPUs. For zoned architecture neutral atom devices, the hardware is divided into entanglement, storage, and readout zones based on the laser-addressable range~\cite{bluvsteinquantum2022, bluvsteinlogical2024}. To execute a two-qubit gate, relevant atoms must be moved from a storage zone to an entanglement zone and returned afterward. Since atom transport error is proportional to the distance moved, it would be necessary to design a completely new heuristic, such as a distance-first strategy. This strategy, as its name suggests, would iteratively select atoms with the shortest required movement distance when constructing substructures. All generated SubQPUs could still be abstracted as VQPUs and stored in the same database.

Second, enabling the select-then-compile workflow requires designing corresponding VQPU selection strategies for different hardware platforms. For linear ion-trap and QCCD architectures, fidelity-first and structure-first strategies may still prove effective, but their efficacy must be further validated on the specific hardware. For zoned architecture neutral atom devices, one could prioritize VQPUs that are sorted according to the distance-first strategy. Once a target VQPU is determined, the subsequent prerequisite is to design a hardware-specific transpilation process. Although the transpiler's modular pipeline architecture is conceptually general, the qubit mapping step must be tailored to the specific hardware. Unlike superconducting processors with fixed and limited connectivity, trapped-ion systems offer logical all-to-all connectivity; however, in scalable QCCD structures, interactions between multiple traps rely on physical ion shuttling, which makes minimizing this slow and error-prone movement the primary focus of the mapping strategy~\cite{upadhyay2022shuttle}. For neutral-atom arrays, which achieve dynamically reconfigurable connectivity via optical tweezers, the qubit mapping objective is to efficiently utilize atom movement and maximize parallelism~\cite{lin2025reuse,huang2024zap}.

Looking forward, the architectural principles embodied by QSteed, namely, resource virtualization and the select-then-compile paradigm, offer a promising path toward scalable, hardware-aware quantum compilation across heterogeneous physical platforms. This abstraction layer not only facilitates compiler portability but also creates opportunities for a unified compilation workflow in heterogeneous quantum clusters, a critical step toward the broader practical deployment of quantum computing systems.

\section{Methods}
We compared QSteed with the most widely used quantum compilation software, including Qiskit (v1.2.4)~\cite{Qiskit} and Pytket (v1.33.1)~\cite{Sivarajah2021}. In the Qiskit and Pytket compilation frameworks, the entire quantum chip is required as the input for the compiler, whereas the QSteed compilation framework selects an appropriate VQPU from the resource virtualization database before performing the transpilation. To represent their best-effort performance, we configure Qiskit to its highest optimization level (\textit{optimization\_level}=3), and Pytket follows the compilation flow adopted in MQTbench~\cite{Quetschlich2023mqtbench}. QSteed employs different compilation strategies, denoted as QSteed-\text{$x$\_$y$\_$z$}, where $x$, $y$, and $z$ represent the VQPU selection strategy, the routing heuristic, and the initial layout method, respectively.

We selected algorithmic circuits from the QASMBench suite~\cite{10.1145/3550488} and generated random circuits using gates like CNOT, RZZ, and RX to form our benchmark set. Each algorithmic benchmark was compiled and executed five times, with the median value taken as the performance metric. For random benchmarks, the metric was computed as the average over more than 50 circuit instances. Specifically, we adopted the following evaluation metrics:

\textit{Compilation time.} 
For compilers designed for hardware deployment, shorter compilation times effectively reduce user waiting times and improve hardware utilization.

\textit{Circuit depth.} 
Due to the limited coherence time of qubits in NISQ devices, shallower circuits tend to produce more reliable computational results.

\textit{Gate count.} 
Different quantum processors may employ different native gate sets. Here, we uniformly assume the native gate set is \{CNOT, RX, RY, RZ\}, and we primarily focus on the number of CNOT gates after compilation, as their error rates are typically an order of magnitude higher than those of single-qubit gates.

\textit{Hellinger fidelity.} 
We use the Hellinger fidelity, $F_H$, to quantify the actual fidelity of a compiled circuit. It is defined from the Hellinger distance $d_H$ as~\cite{geng2022hybrid}:
\begin{equation}
F_H = (1-d_H^2)^2 = \left( \sum_{s\in \{0,1\}^n} \sqrt{P_s^{\text{exp}} P_s^{\text{ideal}}} \right)^2,
\end{equation}
where $P^{\text{exp}}$ and $P^{\text{ideal}}$ are the experimental and ideal probability distributions, respectively. This metric intuitively quantifies the deviation between experimental and ideal outcomes. 
All ideal results in this work are obtained using the Qiskit simulator (for circuits with up to 15 qubits)~\cite{Qiskit} or the NVIDIA CUDA-Q simulator (for circuits with more than 15 qubits).

\textit{Circuit cost function.} 
Due to the performance limitations of NISQ hardware, obtaining reliable results for large-scale circuits through hardware sampling becomes extremely challenging. Under such conditions, we adopt a circuit cost function~\cite{kharkov2022arline} as a proxy to evaluate compiler performance. To this end, we assume a simplified noise model: Markovian gate errors without temporal correlations, no crosstalk between parallel operations, decoherence errors approximated as an exponential function of the circuit depth, and the neglect of measurement errors as well as other hardware-specific imperfections. Under these assumptions, the effective fidelity of the circuit can be approximately defined as:
\begin{equation*}
F_{\mathrm{eff}}=K^{D}\prod_{i}F_i^{1q}\prod_{j}F_j^{2q},
\end{equation*}
where $D$ denotes circuit depth, $K$ is a factor that penalizes deep circuits, and $F^{1q}$/$F^{2q}$ represent single/two-qubit gate fidelitie. This metric is similar to the cost function proposed by IBM Quantum in~\cite{Jurcevic2021}. 
To avoid vanishingly small values arising from multiplicative products, we further take the negative logarithm, leading to the circuit cost function $\mathcal{C}$~\cite{kharkov2022arline}:
\begin{equation}
    C = -\log F_{\mathrm{eff}} = -D \log K - \sum_i \log F_i^{1q} - \sum_j \log F_j^{2q}.
    \label{eq-circuit-cost}
\end{equation}
In this work, the fidelity of single-qubit gates is set to the average value, while the fidelity of two-qubit gates is based on their individual values.
Theoretically, $\mathcal{C}$ is strictly negatively correlated with the effective fidelity, which itself serves as a reasonable approximation to the true hardware fidelity under the simplified noise model, capturing the dominant contributions from gate errors and decoherence. Empirical evidence from Supplementary Figure S7 shows that for circuits up to 30 qubits, the circuit cost $\mathcal{C}$ exhibits a clear negative correlation with the Hellinger fidelity $F_H$ (Spearman rank correlation $\rho=-0.686$), demonstrating that circuits with higher $F_H$ generally have lower $\mathcal{C}$. Accordingly, we adopt this metric as a proxy for the expected circuit fidelity in our simulation experiments.


\section*{Acknowledgments}
\subsection*{Funding}
This work is supported by the Beijing Natural Science Foundation (Grant No.Z220002), the National Natural Science Foundation of China
(Grant No. 92365111), the Innovation Program for Quantum Science and Technology (Grant No. 2021ZD0302400), and the National Natural Science Foundation of China (Grant No.92365206, No.T2121001).

\subsection*{General} 
We acknowledge the contributions to the QSteed code by Hao Chen, Jin-Feng Zeng and Bai-Kang Yuan, as well as the helpful discussions with Yanwu Gu.

\subsection*{Author Contributions}
D.E.L. initialized and supervised the project. H.-Z.X., M.-J.H., and D.E.L.
developed the concept and the structure of the quantum compilation framework, H.-Z.X. developed the software package with the help from M.-J.H., Z.-A.W., Y.-L.F., Y.C., X.Z., J.W., W.-F.Z., and Y.-X.J.. X.C. developed the ErrorGnoMark software for quantum benchmarking. H.-Z.X. and D.E.L. prepared the manuscript. Y.J., H.Y., and H.F. contributed to the discussion and analysis of the results.

\subsection*{Competing interests}
The authors declare that there is no conflict of interest regarding the publication of this article.

\subsection*{Data Availability}
The data supporting the plots within this paper are available from the corresponding author upon reasonable request. The code for QSteed is available at: \textit{https://github.com/BAQIS-Quantum/qsteed}


\appendix

\section*{Appendix}

\renewcommand{\thesubsection}{\Alph{subsection}}

\counterwithin{figure}{subsection}
\counterwithin{table}{subsection}
\counterwithin{equation}{subsection}

\renewcommand{\thefigure}{\thesubsection.\arabic{figure}}
\renewcommand{\thetable}{\thesubsection.\arabic{table}}
\renewcommand{\theequation}{\thesubsection.\arabic{equation}}

\subsection{Interfaces and Use Cases}
To facilitate the deployment of QSteed on quantum hardware or quantum clusters, we provide two primary interfaces: one for backend hardware, enabling quantum resource management, and the other for frontend quantum tasks, supporting quantum compilation.

\subsubsection{Backend Interface}
To facilitate the updating and addition of quantum computing resource information, QSteed provides a unified interface for quantum computing resource virtualization management, \texttt{update\_chip\_api}. By specifying the quantum chip name, \texttt{chip\_name}, and a dictionary containing the quantum chip information, \texttt{chip\_info\_dict}, users can easily add or update quantum chip information in the quantum resource virtualization database. A specific usage example is shown below:

\begin{lstlisting}[language=Python]
from qsteed.apis.resourceDB_api import update_chip_api
import json
chip_name = "example"
chip_file = "chipexample.json"
with open(chip_file, "r") as file:
    chip_info_dict = json.load(file)
update_chip_api(chip_name, chip_info_dict)
\end{lstlisting}

Comprehensive usage examples and deployment instructions for the underlying database can be found on the project's open-source GitHub repository~\cite{QSteed}.

\subsubsection{Compiler Interface}
As middleware, QSteed provides the quantum compiler interface, \texttt{call\_compiler\_api}. Through this interface, user tasks can be sent to the quantum compiler, which returns the compilation information along with the compiled QASM circuit upon completion. The compiled QASM can then be sent to the control system for executing quantum computing tasks. A specific usage example is provided below:

\begin{lstlisting}[language=Python]
from qsteed.apis.compiler_api import call_compiler_api

# Assume you can obtain the user's task information and store it as task_info. 
task_info = {
    "circuit": qasm,    # user's quantum circuit
    "transpile": True,  # True-Perform transpilation
                         # False-Only program verification
    "qpu_name": "example",  # If none, automatically selected
    "qubits_list": None,   # Specify the list of qubits to use
    "optimization_level": 2,  # Preset optimization (0-3)
    "passflow": None, # Custom transpilation Passflow
    "vqpu_preferred": "fidelity"  # VQPU selector preference, "fidelity" or "structure"
}
compiled_results = call_compiler_api(**task_info)
compiled_QASM = compiled_results[0]
qubits_to_cbits = compiled_results[1]
compiled_info = compiled_results[2]
\end{lstlisting}

\subsubsection{Use Cases of Transpiler}
If you are interested in exploring quantum compilation algorithms with QSteed without the need for deployment on quantum hardware, you can focus solely on the quantum transpiler module within QSteed. By customizing the quantum backend, initial model, and passflow, you can use the \texttt{Transpiler} to perform quantum circuit transpilation. The following is an example of the implementation:

\begin{lstlisting}[language=Python]
from qsteed import *

rqc = RandomCircuit(num_qubit=5, gates_number=30)
qc = rqc.random_circuit()

# Backend and initial model settings
basis_gates = ['cx', 'rx', 'ry', 'rz']
coupling_list = [(0, 1, 0.991), (1, 2, 0.976), (2, 3, 0.985), (2, 4, 0.994)]
backend_properties = {
    "name": "ExampleBackend",
    "backend_type": "superconducting",
    "qubits_num": 5,
    "coupling_list": coupling_list,
    "basis_gates": basis_gates,
}
backend_instance = Backend(**backend_properties)
initial_model = Model(backend=backend_instance)

# Predefined transpilation passflow
passes = [
    UnrollTo2Qubit(),
    SabreLayout(heuristic='mixture', max_iterations=3),
    UnrollToBasis(basis_gates=basis_gates),
    GateCombineOptimization(),
    OneQubitGateOptimization(),
    ParaSubstitution()
]  # You can also add your custom pass here.
passflow = PassFlow(passes=passes)

# Perform quantum transpilation
transpiler = Transpiler(passflow, initial_model)
transpiled_circuit = transpiler.transpile(qc)
\end{lstlisting}

Additionally, to facilitate the examination of each pass's execution efficiency and the changes in quantum circuits during the transpilation process, QSteed provides a visualization module for transpilation intermediate results, \texttt{TranspilerVis}. This module adopts a three-stage design:
(1) During the transpilation process, it collects real-time execution data for each pass, such as gate counts, circuit depth, and execution time. (2) The collected data is organized into two levels of information structure: detailed and summarized. (3) An interactive interface is then used to provide multidimensional analytical views, including statistical tables, circuit comparisons, model viewing, etc.

This design not only makes the compilation process more transparent and controllable, aiding developers in debugging and optimizing compilation strategies, but also provides intuitive data visualization to help understand the role and performance characteristics of each pass. This is of significant value in enhancing the efficiency of quantum transpiler development and optimization. An example implementation is shown below:

\begin{lstlisting}[language=Python]
from qsteed.transpiler.transpiler_visualization import TranspilerVis, dynamic_draw

transpiler_vis = TranspilerVis(passflow, initial_model)
transpiled_circuit, info, short_info = transpiler_vis.transpile_vis(qc)
dynamic_draw(info, short_info) 
\end{lstlisting}

\begin{figure*}
\centering
\includegraphics[width=0.99\textwidth]{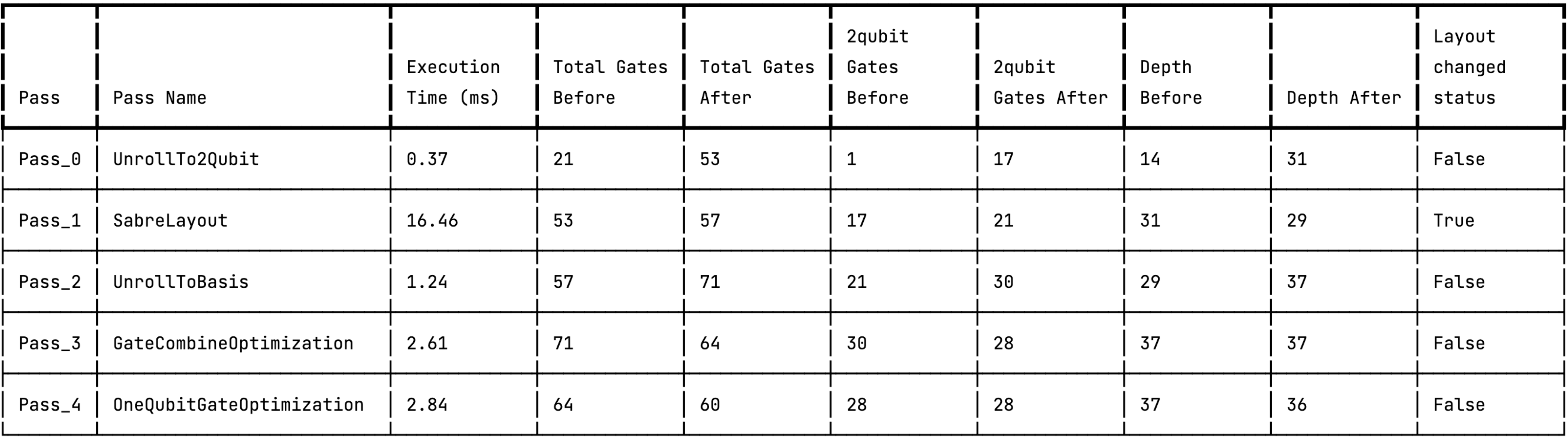}
\caption{A visualization of the results from a circuit compilation process.}
\label{sup-table1}
\end{figure*}

Figure~\ref{sup-table1} illustrates the visualized results of a circuit compilation process. It allows users to easily observe which compilation pass is most dominant and which is the most time-consuming, thereby facilitating targeted performance optimization in future developments.

\subsection{Complexity Analysis of VQPUs Creation}
\subsubsection{Theoretical analysis}
In this subsection, we provide a detailed analysis of the time complexity of the heuristic Algorithm 1 for identifying optimal substructures of the chip.

Let the number of qubits in the chip be $N$, the number of coupling edges be $M$, and the maximum degree be $\Delta$. During the initialization stage (lines 1–6), sorting the $N$ nodes and $M$ edges by fidelity requires $O(N\log N)$ and $O(M\log M)$ time, respectively, while the remaining operations are constant-time, yielding a total cost of $O(N\log N + M\log M)$.

The main loop (lines 7–21) contains three nested loops: (i) the outer loop iterates over target sizes $n=3,\dots,N$; (ii) the method loop iterates over a constant number of strategies; and (iii) the edge loop scans all $M$ seed edges. Inside the \textit{while} loop, the subgraph grows from 2 nodes to $n$ nodes using a priority-queue–based procedure. In each step of the \textit{while} loop, the set of candidate neighbors has size at most $B(n)=\min\{\Delta n, N\}$. Extracting the optimal neighbor from the priority queue requires $O(\log B(n))$, so after $n-2$ iterations the overall cost is $O(n\log B(n))$. In addition, upon generating all candidate subgraphs of size $n$ (approximately $3M$ candidates), the algorithm sorts them by average fidelity for selection (line 20), introducing an extra cost of $O(M\log M)$ for each $n$. Therefore, the per-$n$ complexity is
$$
T_{\text{per-}n} = O(M\,n \log B(n)) + O(M \log M).
$$
Summing over $n=3$ to $N$, we obtain
$$
T_{\text{main}} = \sum_{n=3}^{N} O(M\,n\log B(n)) + \sum_{n=3}^{N} O(M\log M).
$$
The total time complexity of the algorithm is thus
$$
T_{\text{total}} = T_{\text{main}} + O(N\log N + M\log M).
$$

For sparse coupling graphs ($\Delta = O(1)$, as commonly found in superconducting chips), we have $B = \Delta n = O(n)$, then,
\begin{equation*}
\begin{split}
    T_{\text{total}} &= O(MN^{2}\log N) + O(NM\log M) \\
    &\quad + O(N\log N + M\log M).
\end{split}
\end{equation*}
In this case, the first term is dominant, and the overall time complexity can be expressed as 
$$
T_{\text{total}} = O(MN^{2}\log N).
$$
For sparsely coupled superconducting chips, we further have $M = O(N)$, and the complexity simplifies to 
$$
T_{\text{total}} = O(N^{3}\log N).
$$

\subsubsection{Numerical Simulation}
\begin{figure*}
\centering
\includegraphics[width=0.99\textwidth]{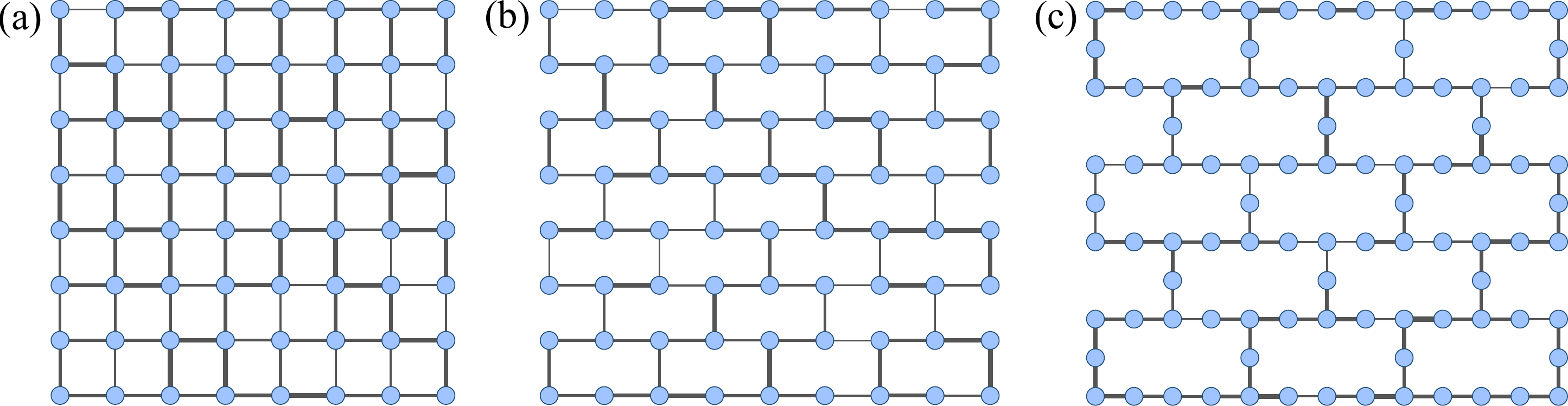}
\caption{Chip topology structures: (a) square lattice, (b) hexagonal lattice, and (c) heavy-hexagonal lattice.}
\label{sup-chips}
\end{figure*}

\begin{figure*}
\centering
\includegraphics[width=0.99\textwidth]{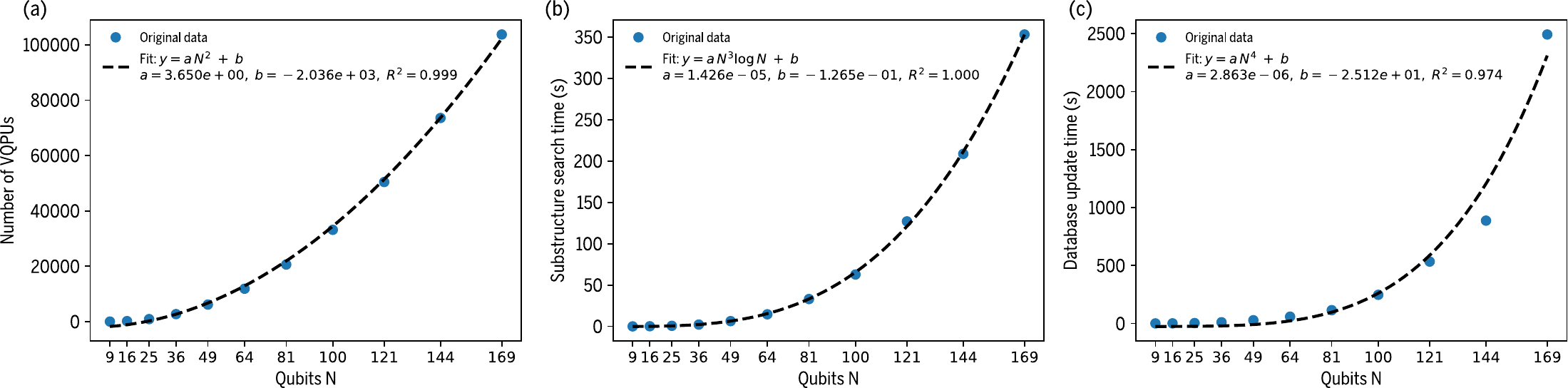}
\caption{Chip with square lattice topology. Panels (a), (b), and (c) respectively show the variation of the number of VQPUs, the time required for substructure identification, and the update time of the resource virtualization database as the number of physical qubits $N$ increases.}
\label{sup-square-time}
\end{figure*}

\begin{figure*}
\centering
\includegraphics[width=0.99\textwidth]{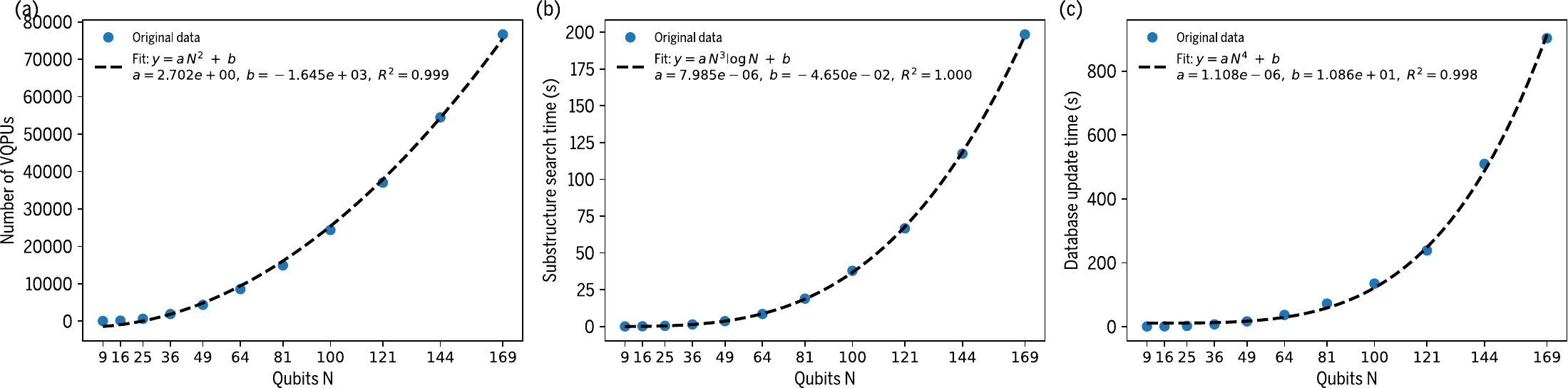}
\caption{Chip with hexagonal lattice topology. Panels (a), (b), and (c) respectively show the variation of the number of VQPUs, the time required for substructure identification, and the update time of the resource virtualization database as the number of physical qubits $N$ increases.}
\label{sup-hex-time}
\end{figure*}

\begin{figure*}
\centering
\includegraphics[width=0.99\textwidth]{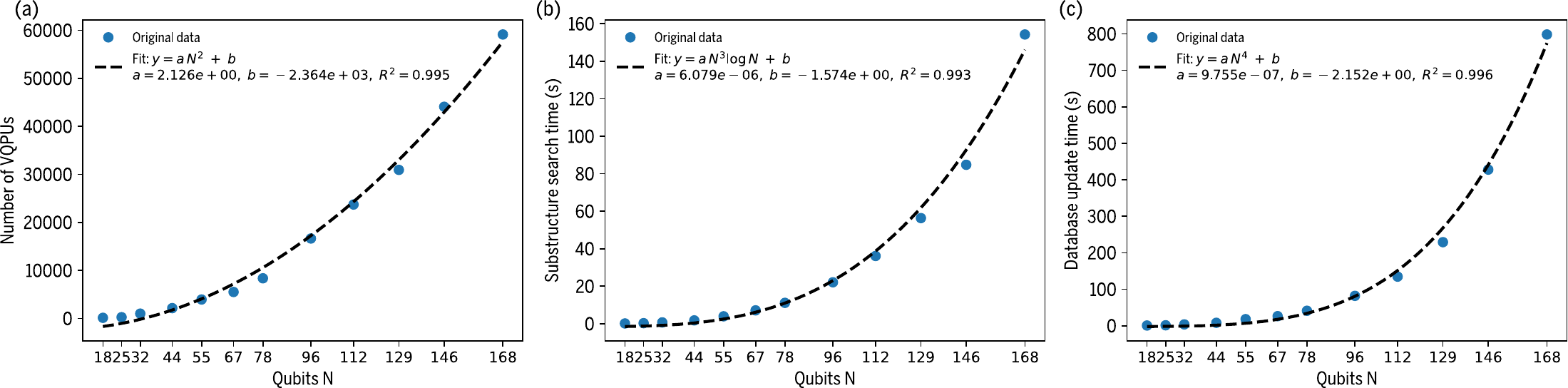}
\caption{Chip with heavy-hexagonal lattice topology. Panels (a), (b), and (c) respectively show the variation of the number of VQPUs, the time required for substructure identification, and the update time of the resource virtualization database as the number of physical qubits $N$ increases.}
\label{sup-heavyhex-time}
\end{figure*}

This section presents the numerical results of the complexity analysis. We consider three chip topologies—square lattice,heavy-hexagonal, and hexagonal—as illustrated in Figure~\ref{sup-chips}. For each structure, we plot the variation of the number of VQPUs, the time required to identify substructures, and the update time of the resource virtualization database as the number of physical qubits increases (see Figure~\ref{sup-square-time}, \ref{sup-hex-time}, and \ref{sup-heavyhex-time}).
It can be observed that the time required by Algorithm 1 for identifying the optimal substructures scales as $O(N^3 \log N)$, which is consistent with the theoretical analysis. The current bottleneck lies in the database update and write operations, with the fitted runtime scaling approximately as $O(N^4)$. The primary reason is that the present implementation uses MySQL with a serial procedure for deleting old records and inserting new ones. Future improvements may include adopting a primary-key–based strategy of differential deletion and batch insertion, as well as redesigning the database architecture or migrating to more efficient systems such as MongoDB or Redis.

\subsection{Weisfeiler-Lehman (WL) subtree kernel algorithm}
The WL subtree kernel measures the similarity between two graphs, $G_1$ and $G_2$, by iteratively updating node labels and computing kernel values. The overall kernel value is given by~\cite{wlKernel}:
\begin{equation}
    K(G_1, G_2) = \sum_{t=1}^{T} k_t(G_1, G_2),
\end{equation}
where $k_t(G_1, G_2)$ represents the subtree kernel value at iteration $t$. At each iteration, this value is computed by comparing the edges of $G_1$ and $G_2$, considering both structural equivalence and, optionally, edge weights. The kernel at iteration $t$ is defined as:
\begin{equation}
    k_t(G_1, G_2) = \sum_{(u, v) \in E_1} \sum_{(x, y) \in E_2} \delta((u, v), (x, y)) \cdot f(w_{uv}, w_{xy}),
\end{equation}
where $\delta((u, v), (x, y)) = 1$ if $u = x$ and $v = y$; otherwise, $\delta = 0$. The function $f(w_{uv}, w_{xy})$ evaluates as $\frac{1}{\exp((w_{uv} - w_{xy})^2)}$ when edge weights are considered, or $f(w_{uv}, w_{xy}) = 1$ otherwise. At the same time, node labels are updated by aggregating the current label of the node with the labels of its neighbors, generating a unique new label:
\begin{equation}
    l_t(v) \leftarrow \text{unique}\left(l_{t-1}(v) \, || \, \text{sorted}(\{ l_{t-1}(u) \mid u \in \mathcal{N}(v) \})\right),
\end{equation}
where $l_t(v)$ denotes the label of node $v$ at iteration $t$, while $\text{unique}$ represents a mapping function that ensures label uniqueness, and $||$ denotes the concatenation operator. A higher kernel value $K(G_1, G_2)$ indicates greater structural similarity between the two graphs. By computing the kernel values between the quantum circuit and all available VQPUs, the module identifies the VQPU with the highest structural resemblance. In cases where multiple VQPUs yield the same kernel value, the one with the highest overall fidelity is selected.

\subsection{Components of the pass module}
The \texttt{Pass} module comprises multiple core components responsible for critical functions. These include the decomposition of multi-qubit gates, the expansion of equivalent quantum gates, qubit mapping and routing, quantum circuit optimization, and the tuning of parameters within parameterized quantum circuits. Our implementation of qubit mapping and routing was described in the main body of this paper. In this section, we provide a detailed exposition of the other aforementioned components.

\subsubsection{Decomposition} 
Numerous theoretical methods have been proposed for universal quantum gate decomposition, including QR decomposition~\cite{mottonen12006decompositions}, quantum Shannon decomposition~\cite{1629135}, and cosine-sine decomposition (CSD)~\cite{PhysRevLett.93.130502}.
In addition, approximate quantum circuit synthesis algorithms have been shown to significantly reduce gate counts~\cite{9259942,10.1145/3505181}; however, their exponential computational complexity limits their practical scalability to small circuits involving only 2–5 qubits.
Reinforcement learning–based approaches have also demonstrated substantial improvements in gate efficiency~\cite{PhysRevLett.125.170501,moro2021quantum}, but their applicability remains largely constrained to 1–3 qubit gates due to the high cost of training.

This study does not aim to design new or more efficient synthesis techniques.
Instead, to decompose an arbitrary $n$-qubit gate into a set of universal basic gates, the current version of QSteed adopts a combination of well-established techniques: ZYZ decomposition~\cite{PhysRevA.52.3457} for single-qubit operations, KAK decomposition~\cite{PhysRevA.69.032315} for two-qubit operations, and cosine-sine decomposition (CSD)~\cite{PhysRevLett.93.130502} for general multi-qubit operations.

\subsubsection{Unroll} 
This component involves a series of equivalent transformations of quantum gates, primarily aiming to expand quantum gates in the circuit using a predefined set of basic gates. For example, $\text{CZ}(i,j)\equiv\text{H}(j)\text{CNOT}(i,j)\text{H}(j)$ and $\text{SWAP}(i,j)\equiv\text{CNOT}(i,j)\text{CNOT}(j,i)\text{CNOT}(i,j)$.

\subsubsection{Optimization} 
On NISQ hardware, the goal of quantum circuit optimization is to reduce circuit depth and gate count as much as possible while preserving circuit equivalence, thereby improving overall circuit fidelity.
A variety of optimization techniques have been proposed, including pattern-matching methods~\cite{10.1145/3498325}, commutation-based approaches~\cite{ITOKO202043}, machine learning techniques~\cite{li2024quarl}, and algorithm-specific optimization strategies~\cite{10.1145/3470496.3527394,xu2024quafu}.

However, quantum circuit optimization is not the primary focus of this study.
In the current version of QSteed, we adopt only a few basic optimization techniques, such as merging parameterized quantum gates and eliminating redundant gate operations.
For instance, two adjacent CNOT gates can cancel each other out, i.e., $\text{CNOT}(i, j)\text{CNOT}(i, j) = \text{I}$.
Nevertheless, QSteed still achieves excellent compilation performance through the combination of a preconstructed VQPU database and a select-then-compile workflow. Users can further implement advanced optimization components by inheriting from the \texttt{BasePass} class.

\subsubsection{Parameter Tuning} 
This component specializes in compiling parameterized quantum circuits for variational algorithms like VQE and QAOA. By integrating Pyquafu’s auto-differentiation framework~\cite{quafu}, it seamlessly handles hybrid circuits containing both variational (e.g., $\text{RZZ}(i, j, \theta)$) and fixed parameters (e.g., $\text{RZZ}(i, j, \pi)$). The compiler automatically optimizes parameterized gate sequences through algebraic simplification, such as reducing $\text{RZZ}(i, j, \theta) \, \text{RZZ}(i, j, \pi)$ to $\text{CNOT}(i, j) \, \text{RZ}(j, \theta + \pi) \, \text{CNOT}(i, j)$, ensuring efficient circuit execution while preserving variational semantics.

\subsection{Performance Comparisons}
This section provides further details on the compilation performance comparison among QSteed, Qiskit, and Pytket. All classical computations were carried out on a workstation equipped with an 8-core Apple M1 Pro CPU and 16 GB of memory. QSteed implements a diverse array of compilation strategies, which are denoted as QSteed-\text{$x$\_$y$\_$z$}. In this nomenclature: The variable $x$ represents the VQPU selection strategy. Specifically, \texttt{fid} corresponds to the fidelity-first strategy, while \texttt{struc} represents the structure-first strategy. The variable $y$ designates the heuristic function employed in the SABRE algorithm. The variable $z$ indicates the initial layout selection for the SABRE algorithm. Here, \texttt{rand} stands for random initialization, \texttt{degree} represents degree initialization, and \texttt{weight} denotes weight initialization. QSteed was originally implemented entirely in Python. Because the qubit-mapping and DAG passes were among the more time-consuming components of the system, we subsequently re-implemented them in C++ to improve performance. In the subsequent descriptions of QSteed compilation strategies, the suffix \texttt{Py} refers to results obtained with the pure-Python implementation, whereas \texttt{C++} denotes runs in which the critical passes are accelerated by C++.

\begin{figure*}
\centering
\includegraphics[width=0.99\textwidth]{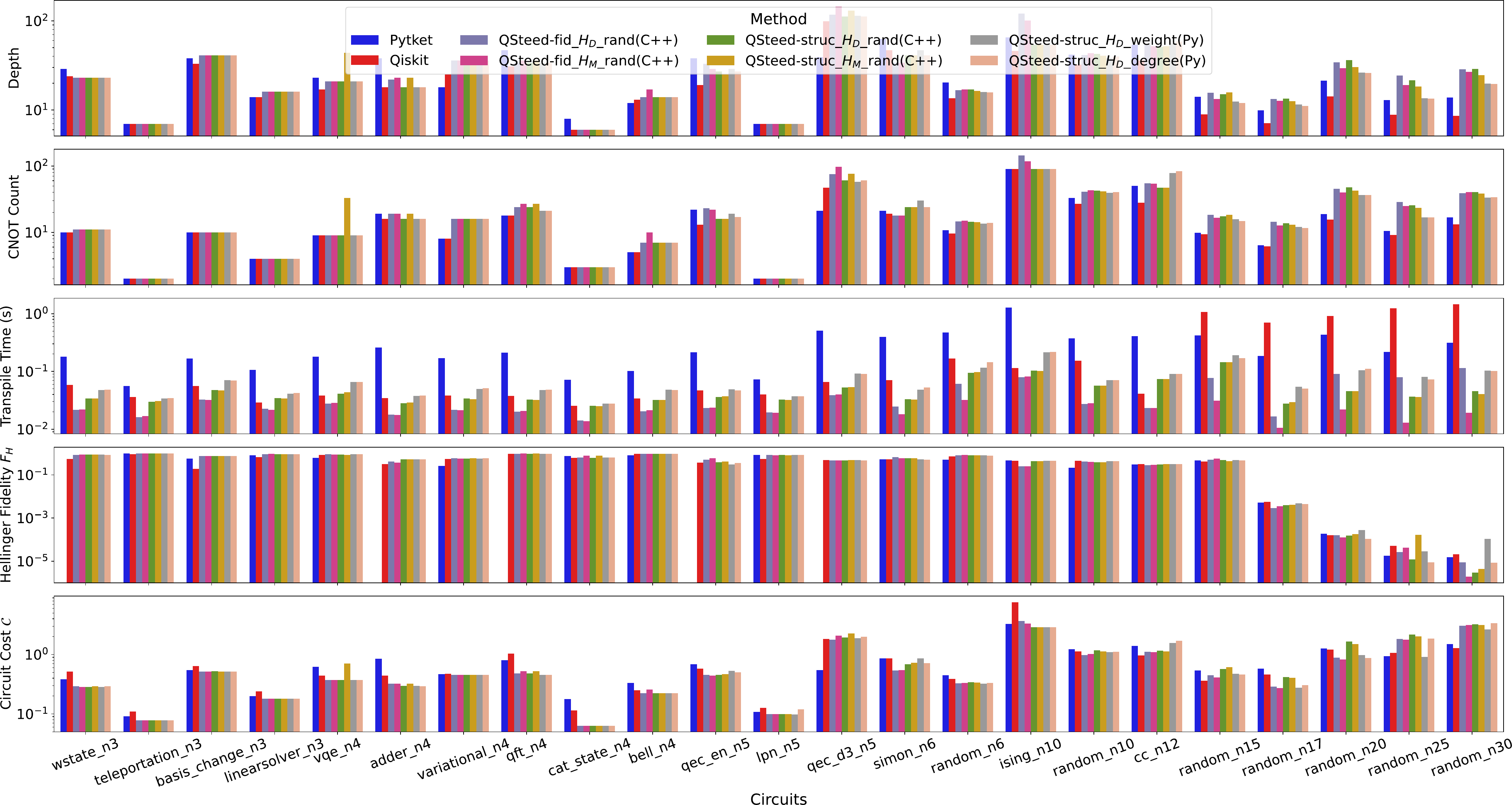}
\caption{Performance comparison of different quantum translators on the Baihua quantum processor. The results for each benchmark circuit represent the median of 5 runs, while the results for random types are averaged over more than 50 runs of randomly generated circuits. The vertical axes, from top to bottom, represent the compiled circuit depth, the number of two-qubit CNOT gates, the transpilation time, the Hellinger fidelity $F_H$ (higher is better), and the circuit cost function $\mathcal{C}$ (lower is better). Except for the random circuits, the circuit names on the horizontal axis are derived from the QASMBench benchmarking suite~\cite{10.1145/3550488}.
}
\label{sup-fig-small-baihua}
\end{figure*}

\begin{figure*}
\centering
\includegraphics[width=0.99\textwidth]{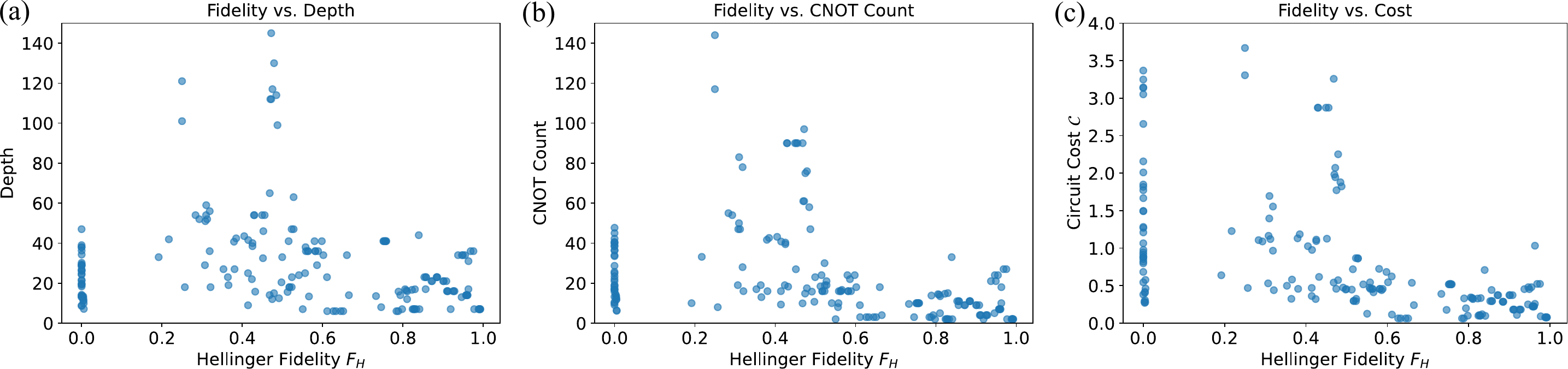}
\caption{Negative correlation between Hellinger fidelity and circuit-complexity metrics. Scatter plots show the relationship between Hellinger fidelity $F_{H}$ and (a) circuit depth, (b) two-qubit (CNOT) gate count, and (c) the composite circuit cost for the same dataset used in Figure~\ref{sup-fig-small-baihua}. All three panels exhibit a statistically significant negative monotonic trend: Spearman rank correlations are $\rho_{\text{depth}} = -0.288$, $\rho_{\text{count}} = -0.576$, and $\rho_{\text{cost}} = -0.686$, with $p < 10^{-4}$ in every case. These results indicate that higher-fidelity circuits tend to be shallower, require fewer CNOT gates, and incur lower circuit cost. Because the circuit cost metric $\mathcal{C}$ yields the strongest (most negative) correlation with fidelity, we adopt $\mathcal{C}$ as a proxy for the expected circuit fidelity in the main text.
}
\label{sup-fig-correlation}
\end{figure*}

\begin{figure*}
\centering
\includegraphics[width=0.98\textwidth]{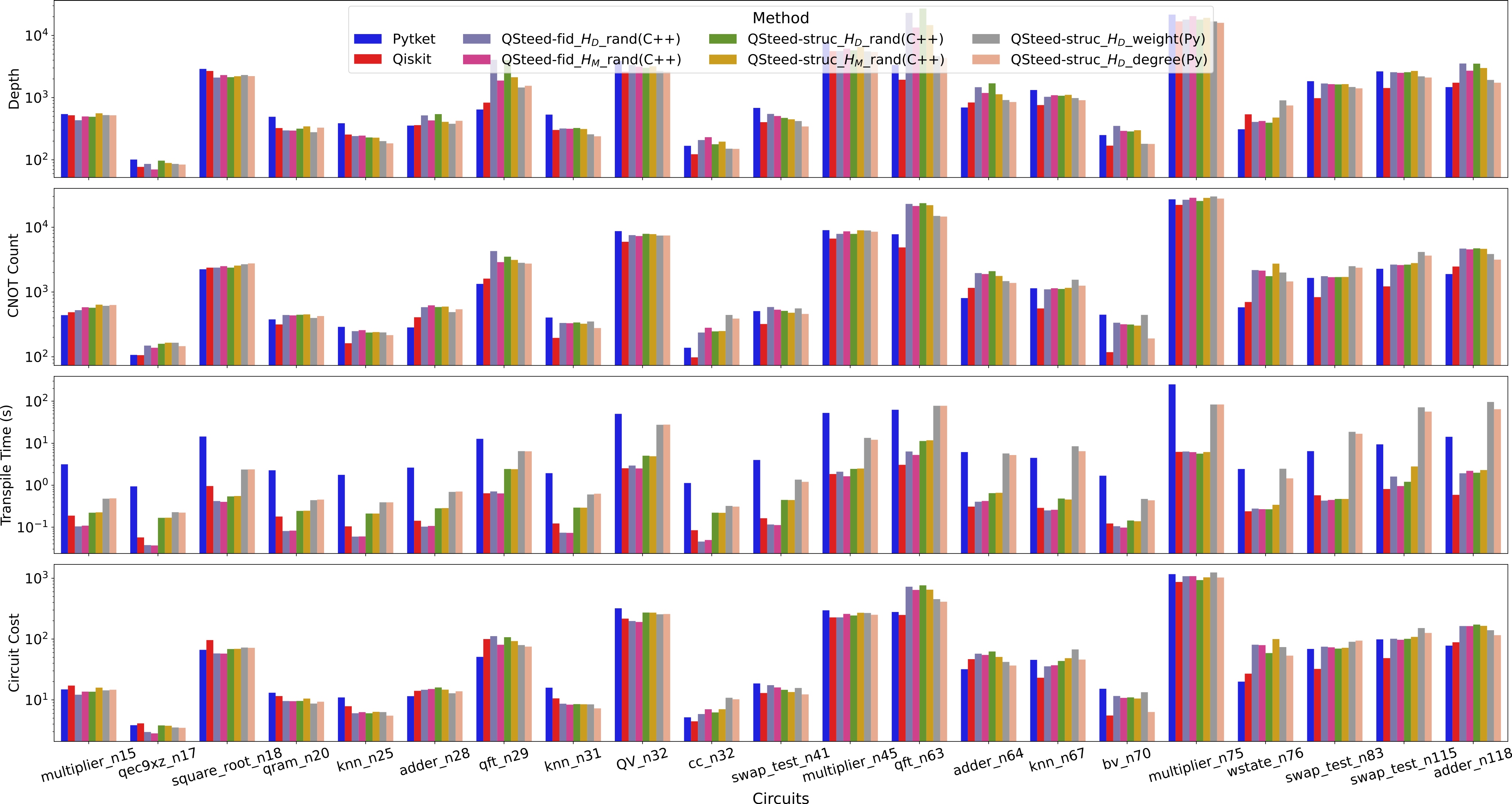}
\caption{Due to the impact of quantum hardware noise, the sampling results of large-scale quantum circuits are often unreliable. Moreover, obtaining ideal computational results for such circuits from simulators remains challenging, making the calculation of $F_H$ infeasible. Therefore, instead of running the circuits compiled by the three compilers on real quantum machines, we simulated and calculated the fidelity of each benchmark circuit based on the parameter information of the Baihua chip. The results for each benchmark circuit represent the median of 5 runs. The vertical axes, from top to bottom, represent the compiled circuit depth, the number of two-qubit CNOT gates, the transpilation time, and the circuit cost function $\mathcal{C}$ (lower is better). The circuit names (trailing number denotes qubits) on the horizontal axis are derived from the QASMBench benchmarking suite~\cite{10.1145/3550488}.}
\label{sup-fig-big-baihua}
\end{figure*}

\begin{figure*}
\centering
\includegraphics[width=0.98\textwidth]{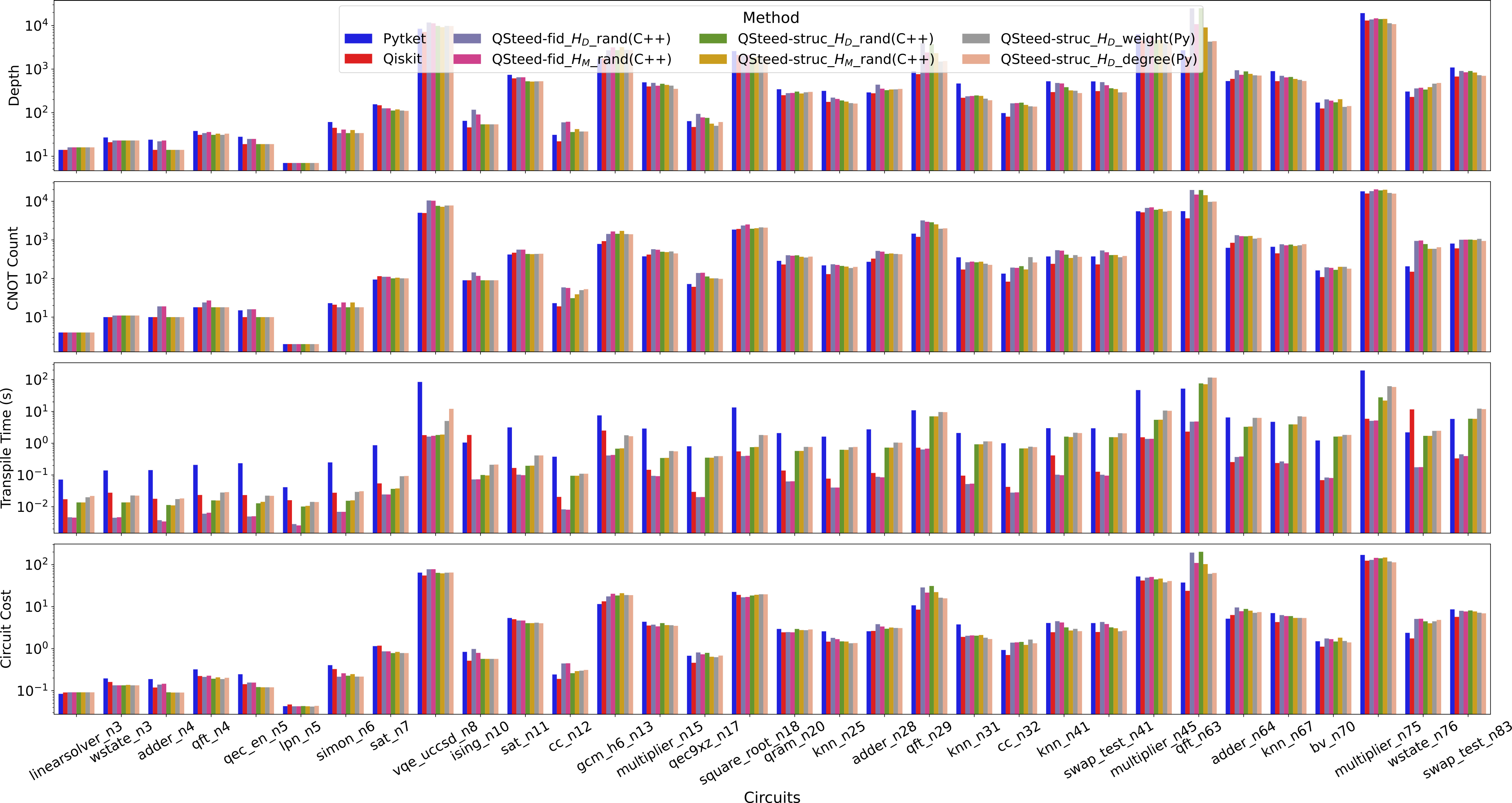}
\caption{Performance comparison of various quantum translators through the emulation of the Willow quantum processor. The chip parameter information is sourced from~\cite{google2024}. The results for each benchmark circuit represent the median of 5 runs. The vertical axes, from top to bottom, represent the compiled circuit depth, the number of two-qubit CNOT gates, the transpilation time, and the circuit cost function $\mathcal{C}$ (lower is better). The circuit names (trailing number denotes qubits) on the horizontal axis are derived from the QASMBench benchmarking suite~\cite{10.1145/3550488}.}
\label{sup-fig-Willow}
\end{figure*}

\subsubsection{Experimental Results and Fidelity–Cost Relationship}
Figure~\ref{sup-fig-small-baihua} presents the benchmarking results obtained on the Baihua processor. We first retrieved the latest calibration data for the chip from the cloud and verified its operational status using ErrorGnoMark~\cite{ErrorGnoMark}, confirming that it was functioning within a reasonable regime. Based on the calibration data, we constructed a VQPU database locally on a personal computer. Each benchmark circuit was then compiled using different toolchains, and the resulting circuits were executed on the cloud via \texttt{QuarkStudio}~\cite{QuarkStudio} to obtain real hardware sampling results, from which Hellinger fidelity $F_{H}$ was computed. We also computed the cost $\mathcal{C}$ using the same set of circuits. All results are summarized in Figure~\ref{sup-fig-small-baihua}.

Figure~\ref{sup-fig-correlation} visualizes the correlations between circuit fidelity and key compilation metrics using scatter plots. The plots show that circuit depth, CNOT gate count, and overall circuit cost $\mathcal{C}$ are all negatively correlated with Hellinger fidelity $F_{H}$. To quantify the strength of these relationships, we compute the Spearman rank correlation coefficients, which are $\rho_{\text{depth}} = -0.546$, $\rho_{\text{count}} = -0.633$, and $\rho_{\text{cost}} = -0.678$, with $p < 10^{-4}$ in all cases. These results indicate that higher-fidelity circuits tend to be shallower, require fewer CNOT gates, and incur lower circuit cost. Among the three metrics, the circuit cost $\mathcal{C}$ exhibits the strongest (most negative) correlation with fidelity, which supports our use of $\mathcal{C}$ as a surrogate predictor of circuit fidelity throughout the main text.

\subsubsection{Simulation results}
Figure\ref{sup-fig-big-baihua} and Figure\ref{sup-fig-Willow} show simulation results on the Baihua and Google's Willow processors, respectively, as supplements to the main discussion.

\begin{figure*}
\centering
\includegraphics[width=0.99\textwidth]{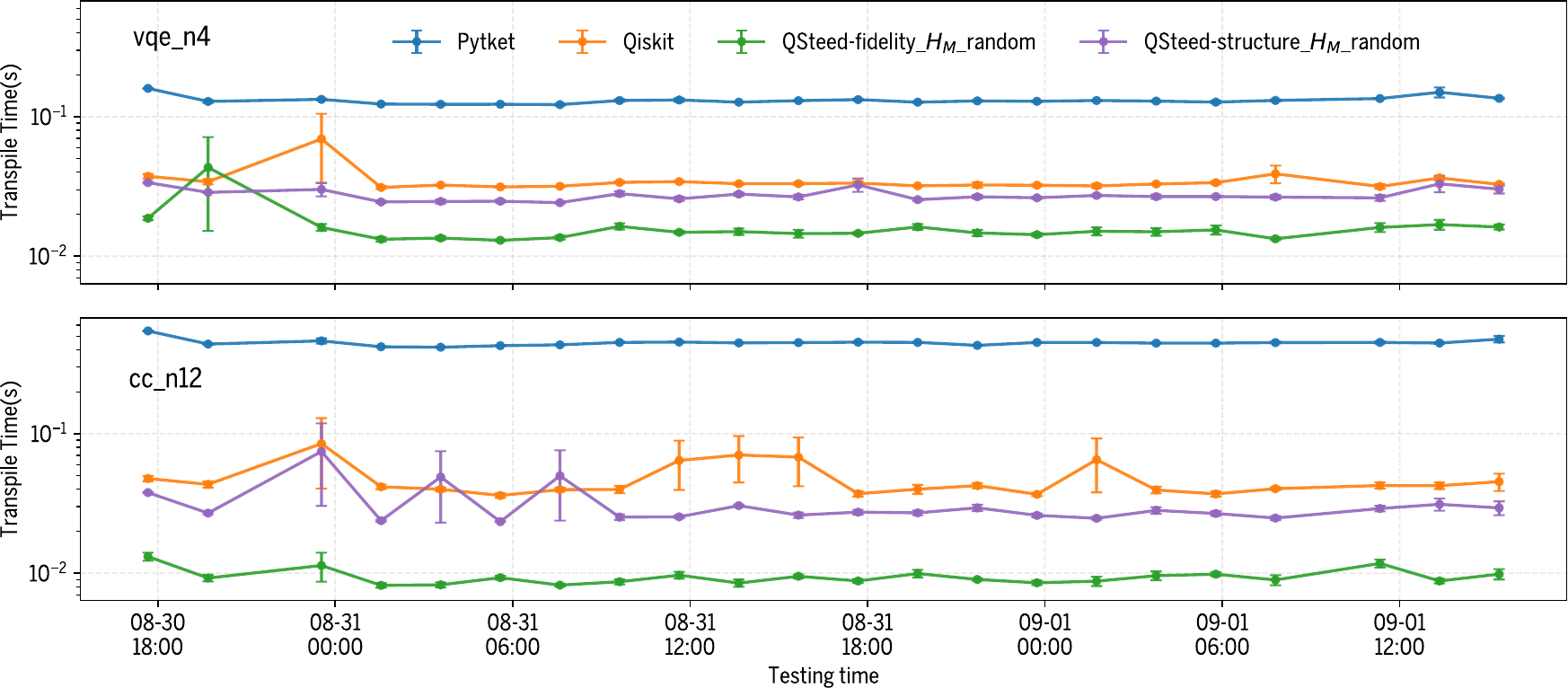}
\caption{Between August 30 and September 1, 2025, circuit tasks were executed every two hours to monitor variations in compilation time. The chip is typically recalibrated once per day.}
\label{sup-fig-timetime}
\end{figure*}

\begin{figure*}
\centering
\includegraphics[width=0.99\textwidth]{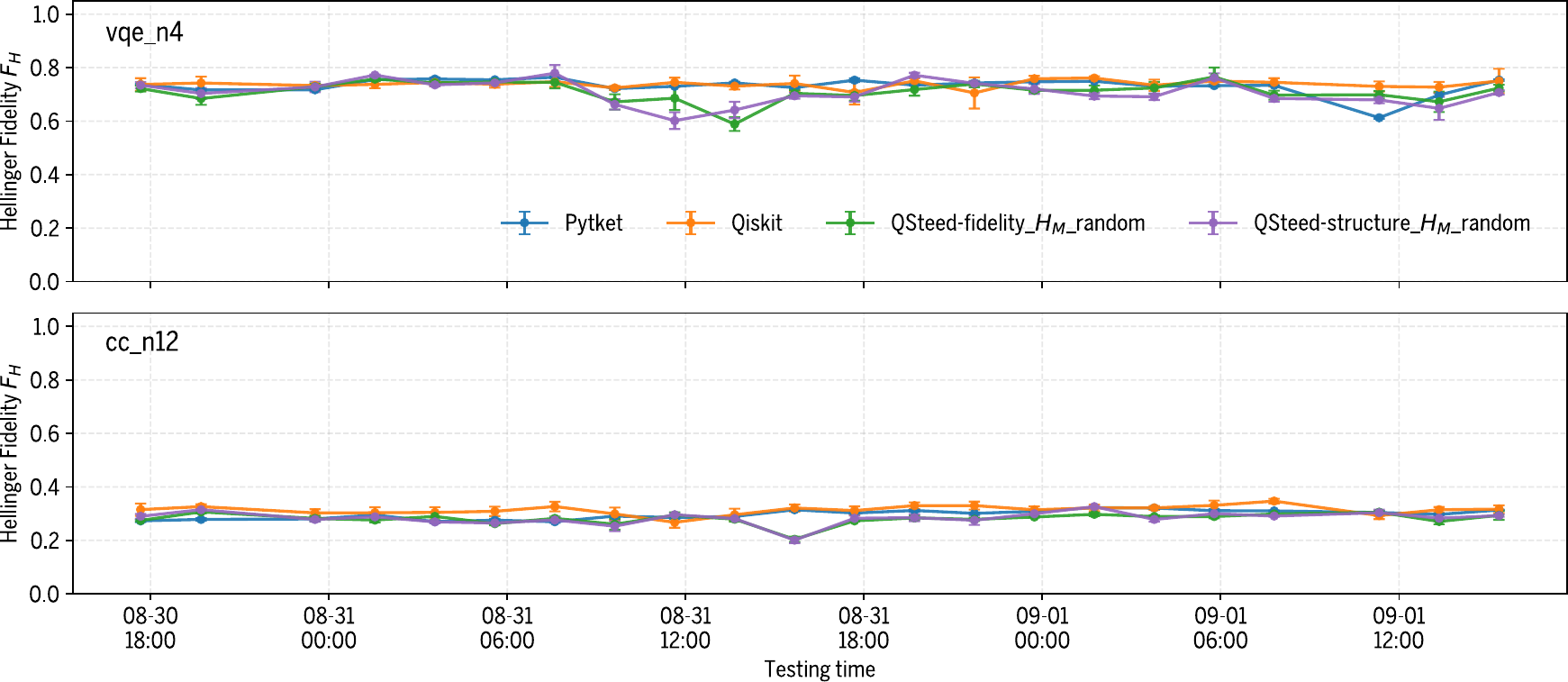}
\caption{Between August 30 and September 1, 2025, circuit tasks were executed every two hours to monitor variations in compiled circuit execution fidelity. The chip is typically recalibrated once per day.}
\label{sup-fig-timefid}
\end{figure*}

\subsubsection{System Stability under Hardware Calibration Dynamics}
Due to environmental factors, quantum hardware may experience calibration parameter drift. To evaluate the robustness of our compilation system under such conditions, we selected the Baihua processor as the target backend (which is typically recalibrated once per day) and developed an automated script that executed tasks every two hours over two consecutive days. Specifically, we compiled and executed the \texttt{vqe\_n4} and \texttt{cc\_n4} circuits every two hours to monitor the stability of the virtualization strategy. As shown in Figure~\ref{sup-fig-timetime} and \ref{sup-fig-timefid}, the compilation performance remained nearly constant during this period.

The VQPU database in QSteed is dynamically linked to the hardware calibration data. When the backend hardware undergoes recalibration, the updated parameters are pushed to the virtualization manager, which refreshes the VQPU database accordingly. It is important to note, however, that while the VQPU database is dynamically synchronized with calibration data, the system cannot detect drift if the hardware itself is not recalibrated in time. This represents a current limitation of our approach. Nevertheless, such issues cannot be fully addressed at the compilation level alone and are beyond the scope of this study. Future improvements could be achieved by integrating the compiler with chip benchmarking and automated calibration systems. For example, fast benchmarking routines (e.g., parallel measurements of GHZ states on two-qubit gates to monitor qubit stability) could be executed periodically, and once qubit drift is detected, an automatic calibration procedure could be triggered to update the hardware parameters.

\bibliography{refClean}

\end{document}